\documentclass[prd,onecolumn, nofootinbib, 11pt]{revtex4}  
\usepackage{graphicx}
\usepackage{amsmath,braket}
\usepackage{amsfonts,mathrsfs}
\usepackage{amssymb}
\usepackage{bm}
\usepackage{appendix}
\usepackage{mathtools}
\usepackage{comment}
\usepackage{bbold}
\usepackage{color}
\usepackage{slashed}
\usepackage[hyperindex=true, citecolor=green]{hyperref}
\usepackage{subfigure}
\usepackage{setspace}
\usepackage{enumitem}
\usepackage{longtable}
\usepackage{wasysym}
\usepackage[usenames,dvipsnames]{xcolor}
\usepackage{bm}
\usepackage{multirow}
\usepackage{changepage}
\usepackage{kantlipsum}
\usepackage{mathtools}
\usepackage[letterpaper, margin=.8in, top=0.85in, bottom=0.85in]{geometry}

\usepackage{tikz}
\usetikzlibrary{decorations.pathmorphing,arrows.meta,bending}

\pdfoutput=1

\def\abs#1{\left | #1 \right | }

\def\osx{\mathcal{O}_{S \to X}}

\def\nn{ \nonumber \\ }

\begin{document}

\singlespacing

\title{Neutrino Oscillation Measurements Computed in Quantum Field Theory}
\author{Andrew Kobach}
\affiliation{Physics Department, University of California, San Diego, La Jolla, CA 92093, USA}
\author{Aneesh V.~Manohar}
\affiliation{Physics Department, University of California, San Diego, La Jolla, CA 92093, USA}
\author{John McGreevy}
\affiliation{Physics Department, University of California, San Diego, La Jolla, CA 92093, USA}

\date{\today}

\begin{abstract}
We perform a calculation in quantum field theory of neutrino oscillation probabilities, where we include simultaneously the source, detector, and neutrino fields in the Hamiltonian.  Within the appropriate limits associated with current neutrino oscillation experiments, we recover the standard oscillation formula.  On the other hand, we find that the dominant contributions to the amplitude are associated with different neutrino mass eigenstates being emitted at different times, such that they arrive at the detector at the same time. This is contrary to the neutrino wave packet picture, where they are emitted simultaneously and  separate as they travel to the detector.  This has direct consequences regarding the mechanisms that lead to a damping of neutrino oscillations for very long baselines. 
Our analysis also provides a pedagogical example of a measurement process in quantum mechanics.
\end{abstract}

\maketitle

\linespread{1}

\section{Introduction}

The experimental observation of flavor transitions in neutrino oscillation experiments implies that neutrinos have mass.  The neutrino field $\nu_\ell$ which couples to the charged lepton $\ell$ is, in general, a linear combination of the fields $\nu_a$ which diagonalize the neutrino mass matrix,
\begin{equation}\label{PMNS}
\nu_\ell = \sum_a U_{\ell a} \, \nu_a\,.
\end{equation}
Global combinations of data from oscillation experiments find good agreement using the following formula for the probability of neutrino oscillation~\cite{Esteban:2016qun, deSalas:2017kay} from flavor $\ell$ to $\ell^\prime$:
\begin{equation}
\label{standardoscprof}
\mathcal{P}^\text{osc}_{\nu_\ell \rightarrow \nu_{\ell'}}(L) \simeq  \left| \sum_{a} U^*_{\ell a} U_{\ell ' a }~e^{-\frac{im_a^2 L}{E_\nu} \xi}  \right|^2 ,
\end{equation}
where $U_{\ell a}$ is the mixing matrix in Eq.~(\ref{PMNS}), $m_a$ is the mass of the neutrino mass eigenstate $\nu_a$,  $L$ is the baseline of the experiment (i.e., the distance between source and detector) and $E_\nu$ is the neutrino energy.  Here, $\xi$ is a constant, which is typically derived to be $\xi = 1/2$, and the neutrino oscillation parameters obtained from oscillation experiments are quoted using this value in Eq.~\eqref{standardoscprof}.

An overview of the standard derivation of the neutrino oscillation phase that leads to $\xi=1/2$, e.g., the one discussed in the PDG~\cite{Patrignani:2016xqp}, can be found in Appendix~\ref{puzzle}.  Here, the oscillation probability $\mathcal{P}(T,L)$ is not only a function of the baseline distance $L$, but also of the transit time $T$ that neutrinos take between their production and detection.  It is then imposed, using an ultra-relativistic limit, that $T\simeq L$ for all neutrino species, and the result is $\xi= 1/2$.  This discussion can be extended to allow for the damping of oscillations for very long baselines $L$.  Many discussions exist in the literature regarding this phenomenon (for example, see Refs.~\cite{Nussinov:1976uw, Kayser:1981ye, Giunti:1991sx, Giunti:2002xg, Giunti:1993se, Grimus:1996av, Kiers:1995zj, Giunti:1997wq, Stockinger:2000sk, Beuthe:2001rc, Beuthe:2002ej, Giunti:2003ax, Bernardini:2004sw, Bernardini:2006ak, Akhmedov:2008jn,  Cohen:2008qb, Akhmedov:2009rb, Naumov:2010um, Akhmedov:2010ms, Akhmedov:2012uu, Jones:2014sfa, An:2016pvi, Boyanovsky:2011xq, Lello:2012gi}). Common to these treatments is that all neutrino mass eigenstates are produced by the source particle in the same region of time.  Under this condition, there is an unavoidable source of damping of oscillations for large enough values of $L$, since, due to their different group velocities, the different mass components of the neutrino wave packet cease to overlap as they undergo their journey from source to detector.  The resulting damping of neutrino oscillations is estimated to be negligible for terrestrial experiments.

There is an inconsistency, however, when computing neutrino oscillations in this manner.  The neutrino transit time $T$ between production and detection is an unmeasured variable in contemporary oscillation experiments.  One can only calculate the probability for the outcome of performed experiments.  If $T$ is unmeasured, one cannot calculate the outcome for such an experiment as a function of $T$.  Instead, all amplitudes with different values of $T$ must be summed.  We can represent this with the following cartoon:
\begin{equation}
\mathcal{A} = \sum_t \sum_a \mathcal{A}(t,m_a) = \sum_{\color{BrickRed} t}  \nonumber
\left(
\scalebox{0.95}{
\begin{tikzpicture}[baseline=(current  bounding  box.center)]
\draw[-{>[scale=1.5,
          length=5,
          width=3]},line width=1.5pt,color=Purple] (0.71,1)--(0.71,2); 
\draw[-{>[scale=1.5,
          length=5,
          width=3]},line width=1.5pt,color=Purple] (4.71,1)--(4.71,2); 
\draw[-{>[scale=1.5,
          length=5,
          width=3]},line width=1.5pt,color=Emerald] (0.71,2)--(0.25,3);       
\draw[-{>[scale=1.5,
          length=5,
          width=3]},line width=1.5pt,color=Green] (4.71,2)--(4.5,3);                 
\draw[-{>[scale=1.5,
          length=5,
          width=3]},line width=1.5pt,color=Cerulean] (0.71,2)--(2,4);  
\draw[line width=1.5pt,color=Cerulean, dashed] (2,4)--(3.29-0.645,6-1);
\draw[-{>[scale=1.5,
          length=5,
          width=3]},line width=1.5pt,color=RoyalBlue] (4.71,2)--(5.75,4);           
\draw[line width=1.5pt,color=RoyalBlue, dashed] (5.75,4)--(6.645,5.7);          
\filldraw[color=Dandelion] (2.645, 5.25) circle (0.1);
\filldraw[color=Dandelion] (6.645, 5.25) circle (0.1);          
\draw[color=Black] (0.5,2) node [align=center] {\Large $\color{BrickRed} t$};
\draw[color=Black] (4.5,2) node [align=center] {\Large $\color{BrickRed} t$};
\draw[color=Cerulean] (1.25,3.3) node [align=center] {$\nu_1$};
\draw[color=RoyalBlue] (5.05,3.3) node [align=center] {$\nu_2$};
\draw[color=Black] (3.1,3) node [align=center] {\LARGE $+$};
\end{tikzpicture}
}
\right)
\end{equation}
Here, the total amplitude $\mathcal{A}$ is the sum over all unmeasured variables, namely the time $t$ that the source particle (depicted by a purple line) decays to a neutrino (blue lines) and other particles (green lines), and the mass of the neutrino $m_a$.  We have drawn two mass eigenstates, i.e., $a=1,2$.  The vertical direction is time, and the horizontal direction is space.  The yellow dot represents the spacetime region of a detector.  As a result of our calculation, we find that the following diagrams are the dominant contribution to the amplitude after integrating over time, i.e., they are the extrema of the path integral:
\begin{equation}
\mathcal{A} \quad \simeq \quad  \nonumber
\scalebox{0.95}{
\begin{tikzpicture}[baseline=(current  bounding  box.center)]
\draw[-{>[scale=1.5,
          length=5,
          width=3]},line width=1.5pt,color=Purple] (0.71,1)--(0.71,2+0.2); 
\draw[-{>[scale=1.5,
          length=5,
          width=3]},line width=1.5pt,color=Purple] (4.71,1)--(4.71,2-0.45); 
\draw[-{>[scale=1.5,
          length=5,
          width=3]},line width=1.5pt,color=Emerald] (0.71,2+0.2)--(0.25,3+0.2);       
\draw[-{>[scale=1.5,
          length=5,
          width=3]},line width=1.5pt,color=Green] (4.71,2-0.45)--(4.5,3-0.45);                 
\draw[-{>[scale=1.5,
          length=5,
          width=3]},line width=1.5pt,color=Cerulean] (0.71,2+0.2)--(2,4+0.2);  
\draw[line width=1.5pt,color=Cerulean, dashed] (2,4+0.2)--(3.29-0.645,6-1+0.2);
\draw[-{>[scale=1.5,
          length=5,
          width=3]},line width=1.5pt,color=RoyalBlue] (4.71,2-0.45)--(5.75,4-0.45);           
\draw[line width=1.5pt,color=RoyalBlue, dashed] (5.75,4-0.45)--(6.645,5.7-0.45);          
\filldraw[color=Dandelion] (2.645, 5.25) circle (0.1);
\filldraw[color=Dandelion] (6.645, 5.25) circle (0.1);          
\draw[color=Cerulean] (1.2,3.3+0.2) node [align=center] {$\nu_1$};
\draw[color=RoyalBlue] (5.05,3.3-0.45) node [align=center] {$\nu_2$};
\draw[color=Black] (3.1,3) node [align=center] {\LARGE $+$};
\end{tikzpicture}
}
\end{equation}
The oscillation probability $\mathcal{P}$ is the square of this amplitude, $\mathcal{P} = |\mathcal{A}|^2$.  Illustrating the result when one calculates this sum over amplitudes is the primary purpose of this paper.  Given the relevant approximations for typical neutrino oscillation experiments, we find that the amplitude is dominated by contributions where the source particle decays at different times such that all the different neutrino species interfere with the detector at the same time.

In this paper, neutrino oscillation experiments will  be treated entirely within quantum field theory, including simultaneously the source, neutrinos, and detector.  To simplify the algebra, the source and detector particles are treated using heavy particle effective theory, where, to lowest order in $1/M$, the particles can be treated as Wilson lines with a constant velocity $v^\mu$. The results we obtain in this paper hold even without this assumption, but the calculations are more involved (they can be found in Appendix~\ref{heavystates}). The main difference from previous analyses is a full quantum mechanical treatment of the problem. We start with the source in an excited state and the detector in the ground state at $t=0$, and compute the probability to find the detector in an excited state at a later time. The source and detector make transitions between their excited and ground states by emitting and absorbing neutrinos. We make no assumptions about the neutrino, or about its journey from source to detector. In quantum mechanics, one cannot make assumptions about the intermediate state, since it is not measured. In the Feynman path integral approach, the neutrino takes all possible journeys between source and detector which have to be summed over, and the different mass eigenstates do not  have to travel together in the path integral. 

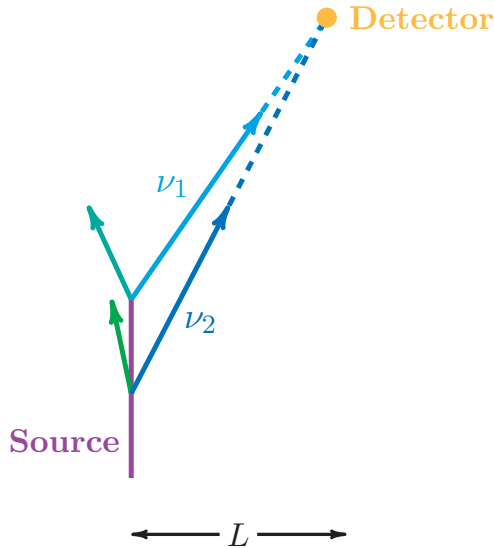
\begin{figure}

\scalebox{1.25}{
\begin{tikzpicture}
\hspace{0.35in} 

\draw[line width=1.5pt,color=Purple] (0.71,0.1)--(0.71,2);
          
\draw[-{>[scale=1.5,
          length=5,
          width=3]},line width=1.5pt,color=Emerald] (0.71,2)--(0.25,3);       
                    
\draw[-{>[scale=1.5,
          length=5,
          width=3]},line width=1.5pt,color=Green] (0.71,1)--(0.5,2);                           
          
\draw[-{>[scale=1.5,
          length=5,
          width=3]},line width=1.5pt,color=Cerulean] (0.71,2)--(2.1,4);  
\draw[line width=1.5pt,color=Cerulean, dashed] (2.1,4)--(2.79,5);

\draw[-{>[scale=1.5,
          length=5,
          width=3]},line width=1.5pt,color=RoyalBlue] (0.71,1)--(1.75,3);           
\draw[line width=1.5pt,color=RoyalBlue, dashed] (1.75,3)--(2.79,5);

\filldraw[color=Dandelion] (2.79,5) circle (0.1);          
          
\draw[color=Cerulean] (1.15,3.2) node [align=center] {$\nu_1$};
\draw[color=RoyalBlue] (1.45,1.75) node [align=center] {$\nu_2$};
\draw[color=Dandelion] (3.8,5) node [align=center] {\small \bf Detector};       
\draw[color=Purple] (0.,0.5) node [align=center] {\small \bf Source};          
\draw[color=Black] (1.85,-0.5) node [align=center] {\small $L$};

\draw[-{>[scale=1,
          length=5,
          width=3]},line width=1pt,color=Black] (1.65,-0.5)--(0.71,-0.5);   
          
\draw[-{>[scale=1,
          length=5,
          width=3]},line width=1pt,color=Black] (2.05,-0.5)--(3.0,-0.5);               

\end{tikzpicture}
}
\caption{[Warning: Do not take the figure literally.  These lines correspond to only the extrema of the path integral.]
A schematic diagram of the dominant contributions to the amplitude for neutrino propagation at oscillation experiments. We show two mass eigenstates $\nu_1$ (light blue) and $\nu_2$ (dark blue), where $\nu_1$ is the lighter neutrino.  The light and dark green lines correspond to other decay products produced when the source decays to the neutrino. The $\nu_1$ and $\nu_2$ contributions to the amplitude have to be added before squaring the amplitude to get the probability. The source (purple lines) is stationary in the figure, as occurs when neutrinos are produced in reactor experiments. This configuration is the sum of dominant amplitudes when $\nu_1$ and $\nu_2$ are emitted from the source at different times, so that they arrive at the detector at the same time.}
\label{spacetimefig} 

\end{figure}

We work in the limits $E_\nu \gg m_a$,  $E_\nu \gg 1/L$, where the detection amplitude can be estimated analytically. The amplitude is dominated by a semiclassical trajectory, which is determined by the structure of the amplitude integral. It is important to keep in mind that this trajectory depends on the entire experiment, i.e., on the properties of the source, detector, and neutrino, and should not be taken too literally. If the source amplitude evolves in time as a pure exponential $e^{-i \Delta_S t - \Gamma_S t/2}$,
we find that the dominant semiclassical trajectory is the one shown in Fig.~\ref{spacetimefig}. The two neutrino mass eigenstates $\nu_{1,2}$ are emitted at different times from the source, in such a way that they arrive at the detector at the same time.  
Figure~\ref{spacetimefig} cannot be taken literally as a spacetime diagram of the neutrino oscillation experiment. It represents the semiclassical trajectory which dominates the amplitude for the calculation that we are doing in this paper. A different calculation with a different time-dependence for the source, such as a source which is turned on and off, requires recomputing the large $L$ amplitude, and the result is not determined by using Fig.~\ref{spacetimefig}.

The outline of this paper is as follows. In Sec.~\ref{heavysources}, we discuss the HQET-based formalism we use for neutrino sources. Section~\ref{stationarysource} computes the oscillation probability for a stationary source and detector treated as two-state systems. Section~\ref{movingsource} generalizes the calculation to a moving source and stationary detector. We show how the picture in Fig.~\ref{spacetimefig} follows from the calculations in Secs.~\ref{stationarysource}, \ref{movingsource}, and discuss the implications of our results in Sec.~\ref{sec:conc}. Appendix~\ref{puzzle} summarizes the factor of two puzzle for $\xi$. Appendix~\ref{heavystates} shows how our results can be generalized, by including localization effects and recoil corrections for the source, and how the results can be applied to sources other than two-state systems. In particular, the results of the appendices show that calculations are applicable for both reactor neutrino experiments, and accelerator experiments where the neutrinos are produced by pion beams.

\section{Heavy Sources and Detectors of Neutrinos}
\label{heavysources}

In oscillation experiments, neutrinos are produced by the decay of a quasi-stable source particle, which is sufficiently localized within a region of space.\footnote{If this were not the case, e.g.,  if the state of the source particle were well approximated instead by a momentum eigenstate, its spatial wave function would be approximately a constant over all space, and it would not be possible to observe oscillations as a function of the experiment's baseline distance~\cite{Cohen:2008qb}.  }  If there exists a reference frame in which the source is non-relativistic, and remains so through time scales of order its lifetime, then such a state can systematically be treated within heavy particle effective field theory.
To illustrate, the amplitude for a free, non-relativistic, scalar particle with mass $M$ to propagate from position eigenstate ${\bf x}$ to ${\bf x}'$ is:
\begin{equation}
\label{heavysource}
\bra{{\bf x}'} e^{-iH_0\tau} \ket{{\bf x}} = \delta^3({\bf x' - x}) e^{-iM\tau} + \mathcal{O}\left( \frac{1}{M} \right)  .
\end{equation}
Here, $\tau$ is the time in the particle's rest frame, and $H_0 \simeq M + {\bf p}^2/2M $.  The limit $M\rightarrow \infty$, i.e., ignoring additional terms in Eq.~\eqref{heavysource}, neglects the spread of the source's spatial wave function in time due to free evolution.  
Therefore, the state can be taken to be arbitrarily localized in the $M\rightarrow \infty$ limit.  A useful mnemonic is:
\begin{equation}
\Delta p \cdot \Delta x \sim 1 ~ \rightarrow ~ \Delta v \cdot \Delta x \sim \mathcal{O}\left(\frac{1}{M}\right) ,
\end{equation} 
so that heavy particles in quantum mechanics can simultaneously have definite position and velocity, but not definite position and momentum~\cite{Jenkins:1990jv}. For the purposes of our present analysis, we will be treating the source as a particle with $\Delta x\sim 0$ and $\Delta v\sim 0$.  Corrections can be included in the $1/M$ expansion, just as in heavy particle effective theory, but such corrections are negligible for terrestrial neutrino experiments.\footnote{Taking an accelerator experiment as an example, the pion mass is $M\sim 100$ MeV, and if the pion's spatial localization is $\Delta x \sim \mathcal{O}(1\, \text{m})$, its momentum uncertainty is $\Delta p \sim 1/\Delta x$.  If so, then an estimate of the ratio of the first and zeroth order contributions to the time evolution of the state, i.e., the ratio of $\Delta p^2\tau/2M$ and $M\tau$,  is $[(\Delta p)^2 /(2M)]/M \sim \mathcal{O}(10^{-30})$.  If the source particle is a radioactive nucleus, confined within a spatial region of scale $10^{-10}$\,m, the diameter of Hydrogen, then $[(\Delta p)^2 /(2M)]/M \sim \mathcal{O}(10^{-17})$.  These ratios are small, and neglecting $1/M$ corrections to Eq.~\eqref{heavysource} is a good approximation. This approximation can break down, however, for extremely long distances of propagation, where the spread of the source's wave function in time leads to a damping of neutrino oscillations, i.e., when $\Delta m^2 L (\Delta p)/p^2$ can become $\mathcal{O}(1)$.  But these propagation distance scales are typically of order $10^{10}$\,m or farther, and it seems somewhat unlikely, given the current state of technology, that in the near future neutrino detectors will be built on one of Jupiter's moons in order to measure the $\mathcal{O}(1/M)$ corrections in Eq.~\eqref{heavysource}. }  

We now explain the Hamiltonian we will use to compute the neutrino oscillation probability. The system consists of a source $S$, a detector $D$ and neutrino fields. The source can transition to a neutrino $\nu$ and a collection of other particles $X$.  In the heavy particle limit, the source travels along a straight worldline~\cite{Schwinger:1951nm, Peskin:1983up, Manohar:2000dt} from which neutrinos and other particles are produced (see Appendix~\ref{heavystates} for more details). The source $S$ behaves like a $B$ meson in HQET~\cite{Manohar:2000dt}. The $S \to \nu X$ decay can be treated like $B \to D$ decays if $X$ is also a heavy state, as in nuclear $\beta$ decay, or like $B \to \pi$ decays if $X$ contains light states, as in $\pi \to \ell \nu_\ell$ decay. In either case, we can still use a $1/M_S$ expansion for the source, which simplifies the integrals that need to be evaluated.

\section{Oscillations with a Stationary Source}
\label{stationarysource}

Here, we consider a scalar theory with a (heavy) stationary source, located at the origin, which can have two quantum states. The source is originally in the excited state $S_E$, which decays to the ground state $S_G$ with no recoil, and creates a quantum of the scalar field $\phi_a$ with mass $m_a$.\footnote{Actual neutrinos are fermions.  However, this complication does not change the oscillation behavior, for the following reason.  The location of the pole in the scalar propagator $G_S$ is the same as for the fermionic propagator $G_F$, since $G_F = (i\gamma \cdot \partial + m)G_S$.  In the limit $E_\nu \gg 1/L$, the pole is the dominant contribution for neutrinos traveling macroscopic distances, i.e., it can be approximated as being on shell, the fermionic version of the calculation differs only by non-exponential prefactors to the oscillation phase. \label{foot:fermions}} $\Delta_S$ is the mass splitting of the two source states. Since the source can decay from $S_E \to S_G + \phi$, interactions of the source with the $\phi$ field produce a width for the source in the excited state.
The well-known Weisskopf-Wigner result~\cite{sakurai1995} shows that this interaction can be included by the replacement $\Delta_S \rightarrow \Delta_S - i\Gamma_S/2$, where $\Gamma_S$ is the decay rate of the source~\cite{Chiu:1977ds,Akhmedov:2008jn, Boyanovsky:2011xq}.

We treat the detector in a similar fashion, where the heavy detector particle (called $D_G$ for ground state), located at position ${\bf L}$, can absorb a field $\phi_a$, and transition to another heavy state (called $D_E$ for excited state) with no recoil, where $\Delta_D$ is the detector mass splitting. This model is similar to the one used in Ref.~\cite{Dickinson:2016oiy}, and is essentially the famous Fermi two-atom problem~\cite{fermi1932}.  See endnote 4 of Ref.~\cite{Dickinson:2016oiy} for the history of this problem. In order to model a detector that can measure more than one energy, many distinguishable two-level systems with different energy splittings $\Delta_{D_j}$ can be placed at the same location. The detector width $\Gamma_D$ can be included in the same way as the source width $\Gamma_S$. We drop it, since it does not lead to any new features in the calculations.

In Appendix~\ref{fullqftrecoil}, we discuss the general calculation, in the full quantum field theory, which leads to the same final result as the one presented in this section, but with considerable increase of complexity.

There are several scalar fields $\phi_a$ with different masses $m_a$. We consider the simple case where the source and detector couple to one linear combination of mass eigenstates $\phi = \sum_a U_a \phi_a $, with $\sum_a |U_a|^2 = 1$. It is trivial to generalize to the case where source and detector couple to different linear combinations, by a  replacement of the mixing matrix $U_a$ with ones for the source and detector. The interaction Hamiltonian is
\begin{equation}
\label{simpleV}
V(t) = \sum_a g \bigg[   U_a^* ~e^{-i\Delta_{S} t}~ \phi^{(-)}_a(t,0) ~\ket{S_G}\bra{S_{E}} +  \sum_{j} U_a~e^{i\Delta_{D_j} t}~\phi^{(+)}_a(t,{\bf L}) ~ \ket{D_{E_j}}\bra{D_G}  \bigg]+ \text{h.c.} ,
\end{equation}
where, 
\begin{eqnarray}
\phi^{(+)}_a(t,{\bf x}) &\equiv& \int \frac{d^3{\bf p}}{(2\pi)^3} \frac{1}{\sqrt{2E^a_{\bf p}}} ~a_{{\bf p},a}~ e^{iE^a_{\bf p}t - i{\bf p \cdot x}}, \\
\phi^{(-)}_a(t,{\bf x}) &\equiv& \int \frac{d^3{\bf p}}{(2\pi)^3} \frac{1}{\sqrt{2E^a_{\bf p}}} ~a^\dagger_{{\bf p},a}~ e^{-iE^a_{\bf p}t + i{\bf p \cdot x}} ,
\end{eqnarray}
and $E^a_{\bf p} \equiv \sqrt{{\bf p}^2 + m_a^2}$.  The source and detector coupling are $g$. We consider sources and detectors which are active for a finite time interval, by including the source and/or detector terms in Eq.~(\ref{simpleV}) only during the active time-window. $\phi^{(+)}$ annihilates $\phi$, and $\phi^{(-)}$ creates $\phi$. The use of only $\phi^{(\pm)}$ in the two terms of $V$ is referred to as the rotating wave approximation. Deviations from this approximation are suppressed due to energy conservation.

The initial quantum state of the system at $t=t_i=0$ is $\ket{i} = \ket{S_E} \otimes \ket{D_G}  \otimes  \ket{0_\phi}$ --- the source is in the excited state, the detector is in the ground state, and there are no $\phi$ quanta excited. We can set up the initial conditions $\ket{i}$, since we have an infinite amount of time before starting the experiment. The initial density matrix is thus
\begin{equation}
\rho_i = \ket{S_E} \bra{S_E}\ \otimes\ \ket{D_G} \bra{D_G}\ \otimes\  \ket{0_\phi} \bra{0_\phi}\,.
\end{equation}
To study whether neutrinos have oscillated, we evolve $\rho_i$ in time using Eq.~(\ref{simpleV}), and trace with the final state density matrix at time $t_f$
\begin{equation}
\label{17}
\rho_f = \mathbb{1}\ \otimes\ \ket{D_E} \bra{D_E}\ \otimes\  \mathbb{1} \,.
\end{equation}
This gives the probability that the detector is in an excited state. It places no restriction on the states of the source and $\phi$ field, which are summed over. In our example, the source and detector are on for long times, and the neutrino time of flight is also large compared to the instrinsic time scales $1/\Delta_{S,D}$. Approximate energy conservation implies that the only source state which contributes is $\ket{S_G}$, and the amplitude of $\ket{S_E}$ is exponentially small. With our Hamiltonian, the only $\phi$ state which contributes is the ground state $\ket{0_\phi}$.\footnote{Calculations using a similar two-state model without the rotating wave approximation, where other $\phi$ states can contribute, are discussed in Ref.~\cite{Dickinson:2016oiy, Fleischhauer}. Within the light cone, 
the answer is found to be the same as in the rotating wave approximation.} Thus we will study the transitions
\begin{equation}
\label{18}
\ket{i} = \ket{S_E} \otimes \ket{D_G}  \otimes  \ket{0_\phi} \quad \to \quad
\ket{f} =  \ket{S_G} \otimes \ket{D_E}  \otimes  \ket{0_\phi}\,.
\end{equation}
If there are several detectors, then all detectors are in the ground state in $\ket{i}$, and we have a set of final states $\ket{f_j}$ where detector $j$ is excited, and all other detectors are in the ground state. Final state configurations where multiple detectors are excited are higher order in the weak interactions, and can be neglected.

The lowest-order contribution in perturbation theory in $g$ that gives rise to the transition between these initial and final states is
\begin{equation}
\label{amp}
\mathcal{A}_j = (-i)^2 \int_{\tau}^{\tau+T} dt_1 \int_0^{t_1} dt_2 ~ \langle f_j | V(t_1) V(t_2) | i \rangle ,
\end{equation}
where the source is on for all time, and the detector is on only during the time window $[\tau, \tau+T]$. $\tau$ is the time of detection, and $T$ is the size of the time bin. This allows us to model an experiment where detector events are counted as a function of time.  A set of such detectors, which are on at different times $[\tau_1, \tau_1+T]$, $[\tau_2,\tau_2+T]$, etc.,\ allows one to compute the detection rate as a function of arrival time at the detector.

To maintain a simpler notation in the intermediate steps, we calculate the transition amplitude in Eq.~\eqref{amp} with $\Delta_S$ as the energy splitting of the source, and use a single $\phi$ mass eigenstate. The full answer is then given by the replacement $\Delta_S \rightarrow \Delta_S - i\Gamma_S/2$ and summing over masses at the end. Here, Eq.~\eqref{amp} becomes
\begin{eqnarray}
\label{simpleamp}
\mathcal{A}_j = -g^2  \int_{\tau}^{\tau+T} dt_1 \int_0^{t_1} dt_2 ~ e^{-i\Delta_{S} t_2 + i \Delta_{D_j} t_1} \langle 0 | \phi^{(+)}(t_1,{\bf L}) \phi^{(-)}(t_2,0) | 0 \rangle ,
\end{eqnarray}
where
\begin{equation}
\langle 0 | \phi^{(+)}(t_1,{\bf L}) \phi^{(-)}(t_2,0) | 0 \rangle = \int \frac{d^3 {\bf p}}{(2\pi)^3} \frac{1}{2E_{\bf p}} e^{-i(t_1 - t_2)E_{\bf p} + i {\bf p \cdot L}} \,.
\end{equation}
Performing the $t_2$ and angular integrals, and changing integration variables to $E_p $: 
\begin{eqnarray}
\label{enint}
\mathcal{A}_j &=& \frac{-ig^2}{(2\pi)^2 L} \int_{\tau}^{\tau+T} dt_1 ~ e^{it_1\Delta_{D_j}} \int_m^\infty dE_p ~ \sin\left(L\sqrt{E_p^2 - m^2} \right) \Bigg[ \frac{e^{-it_1 E_{ p}} - e^{-it_1 \Delta_{S}}}{E_{ p} - \Delta_{S}} \Bigg] \nn
&=&\frac{-ig^2}{(2\pi)^2 L} \int_{\tau}^{\tau+T} dt_1 ~ e^{it_1\Delta_{D_j}} \int_m^\infty dE_p ~
\Biggl[ \frac{e^{iL\sqrt{E_p^2 - m^2}} - e^{-iL\sqrt{E_p^2 - m^2}} } 
{2i} \Biggr] \Bigg[ \frac{e^{-it_1 E_{ p}} - e^{-it_1 \Delta_{S}}}{E_{ p} - \Delta_{S}} \Bigg] .
\end{eqnarray}
The integrand has no poles along the real $E_p$ axis for $E_p > m$, but the individual terms do. We therefore first deform the contour to go slightly above $E_p=\Delta_S$, and then split the integral into four terms given by multiplying out the two factors in square brakets in Eq.~\eqref{enint}.
\begin{figure}
\includegraphics[width=3.25in]{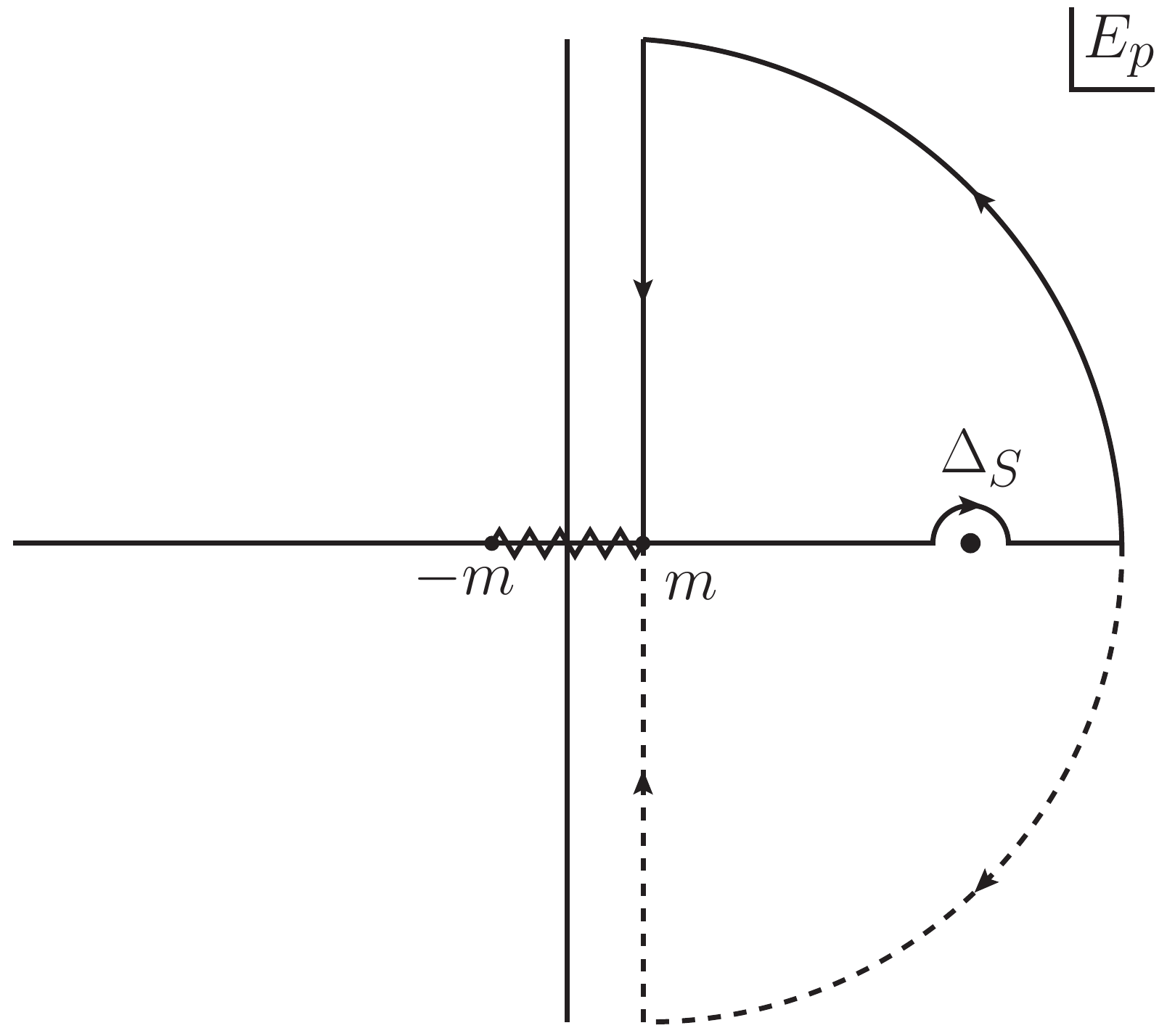}
\caption{The contours used to estimate the value of the energy integral in Eq.~\eqref{enint}.  The vertical parts of the contour are negligible in the limit $\Delta_S \gg 1/L$, $\Delta_S \gg m$, and $\Delta_S \gg 1/|t_1 - L|$. }
\label{contour}
\end{figure}

We utilize residue methods using the contours in Fig.~\ref{contour} to compute the energy integral. Depending on the sign of the exponential for large $E_p$, we consider the closed contour in either the upper or lower half plane, such that the exponential is damped along the circular contour at infinity. The original integral is then given by the residue of the pole at $\Delta_S$ if the contour is closed in the lower half plane, plus the integral along the (dotted) vertical section, or by the (solid) vertical section if the contour is closed in the upper half plane. 
The vertical parts of the contours, parallel to the imaginary axis, are suppressed by powers of $1/L$ in the limit that $\Delta_{S} \gg m$, $\Delta_{S} \gg 1/L$, and $\Delta_{S} \gg 1/|t_1-L|$, which is the case in neutrino oscillation experiments. They give a contribution analogous to near-field effects in classical radiation theory.

If $L>0$ and $t_1 - L < 0 $, i.e., the detector is outside the source's light cone, the integral vanishes
\begin{equation}
\mathcal{A}_j = 0 + \mathcal{O}\left(\frac{1}{L^2}\right)\,,
\end{equation}
up to the power suppressed terms. There cannot be a signal transmitted by the intermediary $\phi$ particle faster than light.  In our calculation, there is a small power suppressed contribution outside the light-cone (from the vertical segment) because of the form Eq.~\eqref{simpleV} where the $\phi$ field has been split into its creation and annihilation parts. If we had instead used the full $\phi$ field in both terms, causality would be exact, and $A_j=0$ outside the light-cone (see Ref.~\cite{Dickinson:2016oiy, Fleischhauer}).

On the other hand, if $L>0$ and $t_1 - L > 0 $, and the detector is well inside the source's light cone, the integral is
\begin{equation}
\label{aj}
\mathcal{A}_j \simeq  \frac{ig^2}{4\pi L} ~e^{iL\sqrt{\Delta_{S}^2 - m^2}}  \int_{\tau}^{\tau+T} dt_1 ~ e^{it_1 (\Delta_{D_j}-\Delta_{S})} 
= \frac{ig^2}{4\pi L} ~e^{iL\sqrt{\Delta_{S}^2 - m^2}}  e^{i (\tau+T/2) (\Delta_{D_j}-\Delta_{S})} 
\frac{2 \sin \left[(\Delta_{D_j}-\Delta_{S}) T/2\right]}{(\Delta_{D_j}-\Delta_{S})},
\end{equation}
up to power suppressed terms.
Making the replacement $\Delta_S \rightarrow \Delta_S - i\Gamma_S/2$, summing over multiple mass eigenstates $\phi_a$:
\begin{equation}
\label{finalamp}
\mathcal{A}_j \simeq \frac{ig^2}{2\pi L}  \sum_a |U_a|^2~e^{iL\sqrt{(\Delta_{S} - i\Gamma_S/2)^2 - m_a^2} + i (\tau+T/2)(\Delta_{D_j} - \Delta_S + i \Gamma_S/2)}\ \frac{\sin\left[(\Delta_{D_j} - \Delta_S + i\Gamma_S/2)T/2\right]}{\Delta_{D_j} - \Delta_S + i\Gamma_S/2} .
\end{equation}
The total probability for the transition includes a sum over the distinguishable detector systems labelled by $j$:
\begin{equation}
\mathcal{P} = \sum_j |\mathcal{A}_j|^2 = \int d\Delta_{D_j}~ \rho_D(\Delta_{D_j}) ~ |\mathcal{A}_j|^2 ,
\end{equation}
where $\rho_D(\Delta_{D_j})$ is the number density of detector systems with energy splitting $\Delta_{D_j}$.  The sinc function\footnote{The sinc function is defined as sinc($x$) $\equiv \sin(x)/x$. } in Eq.~\eqref{finalamp} should be familiar from the usual quantum mechanical derivation of Fermi's Golden Rule.  If $\Gamma_S=0$, then in the limit $\Delta_S T \to \infty$
\begin{equation}
\frac{\sin^2\left[(\Delta_{D_j} - \Delta_S)T/2 \right]}{(\Delta_{D_j} - \Delta_S)^2 } \to \frac 1 2\pi T \delta(\Delta_{D_j} - \Delta_S) ,
\end{equation}
as in Fermi's Golden rule. The only detectors which respond are those with the same energy splitting as the source, and the detection probability is proportional to the time $T$ that the detector is on. If $\Gamma_S \not =0 $, then in the limit $\Delta_S T \to \infty$, $\Gamma_S T \to \infty$
\begin{equation}
\label{21}
\abs{\frac{\sin\left[(\Delta_{D_j} - \Delta_S + i\Gamma_S/2)T/2\right]}{\Delta_{D_j} - \Delta_S + i\Gamma_S/2} }^2 \to \frac{e^{\Gamma_S T/2}}{\left(\Delta_{D_j} - \Delta_S\right)^2 + \Gamma_S^2/4}\,.
\end{equation}
For a non-zero width, the detector response is given by a Lorentzian with a linewidth $\Gamma_S$, and the total probability is not of order $T$, but of order $1/\Gamma_S$, because that is the lifetime of the source. The $e^{\Gamma_S T/2}$ factor converts $e^{-\Gamma_S (\tau+T/2)}$ in $\abs{A_j}^2$ into $e^{-\Gamma_S \tau}$, where $\tau$ is the time the detector is first turned on.

In the limit that $\Delta_S \gg m_a$, the total probability becomes
\begin{equation}
\label{simpleprob}
\mathcal{P} = \frac{g^4 T }{8\pi L^2}\rho_D(\Delta_S) \Bigg| \sum_a |U_a|^2  ~e^{- \frac{im_a^2L}{2\Delta_S} } \Bigg|^2 
\,,
\end{equation}
if $\Gamma_S=0$, and
\begin{equation}
\label{22}
\mathcal{P} = \frac{ g^4 }{2\pi \Gamma_S L^2} \overline \rho_D \Bigg| \sum_a |U_a|^2  ~e^{- \frac{im_a^2L}{2}\frac{\Delta_S}{\Delta_S^2+\Gamma_S^2} - \frac{\Gamma_S}{2}\left(\tau - L - \frac{m_a^2L}{2(\Delta_S^2+\Gamma_S^2)} \right)} \Bigg|^2,
\end{equation}
if $\Gamma_S \not=0$,
where 
\begin{equation}
\label{23}
\overline \rho_D(\Delta_S) = \frac{\Gamma_S}{2\pi} \int d \Delta\ \rho_D(\Delta) \frac{1}{\left(\Delta - \Delta_S\right)^2 + \Gamma_S^2/4},
\end{equation}
is the number of detectors at energy $\Delta_S$, smeared over a width $\Gamma_S$. Since $\Gamma_S \ll \Delta_S$,
\begin{equation}
\label{24}
\mathcal{P} = \frac{ g^4 }{2\pi \Gamma_S L^2} \overline \rho_D(\Delta_S)  \Bigg| \sum_a |U_a|^2  ~e^{- \frac{im_a^2L}{2\Delta_S} - \frac{\Gamma_S}{2}\left(\tau - L - \frac{m_a^2L}{2\Delta_S^2}\right)} \Bigg|^2\,.
\end{equation}

Eq.~\eqref{simpleprob} contains the well-known oscillation term, the classical $1/L^2$ area law  for flux from a point source in three spatial dimensions, as well as an exponential damping factor, which depends on the masses $m_a$.  To further simplify the expression, it is useful to note that the velocity of an ultra-relativistic particle in the lab frame with energy $E$ can be approximated as
\begin{equation}
\label{relvel}
\frac{1}{v_a} = \frac{E}{\sqrt{E^2 - m^2_a}} \simeq 1 + \frac{m_a^2}{2E^2} .
\end{equation}
Since $\Delta_S$ is the energy of all the intermediary particles $\phi_a$, independent of their masses, the probability Eq.~(\ref{23}) has the simple form
\begin{eqnarray}
\label{eqdifftimes}
\mathcal{P} &\simeq& \frac{ g^4 }{2\pi \Gamma_S L^2} \overline \rho_D \Bigg| \sum_a |U_a|^2  ~e^{- \frac{im_a^2L}{2\Delta_S} - \frac{\Gamma_S}{2}\left(\tau - L/v_a \right)} \Bigg|^2 .
\end{eqnarray}
The prefactor $g^2\overline\rho_D/2\pi\Gamma_S L^2$ is associated with the probability of production, and detection, and the inverse square law. The oscillation factor is
\begin{equation}
\label{oscprob1}
\mathcal{P}_\text{osc} \simeq  \Bigg| \sum_a |U_a|^2  ~e^{- \frac{im_a^2L}{2\Delta_S} - \frac{\Gamma_S}{2}\left(\tau - L/v_a \right)} \Bigg|^2 .
\end{equation}
The derivation of $\mathcal{P}_\text{osc}$ did not assume anything about the journey of the intermediate neutrino state. In particular we made no mention of any neutrino wavepackets travelling from source to detector. The result Eq.~(\ref{oscprob1}) follows by using time-ordered perturbation theory, and keeping the leading terms in the $L \to \infty$ limit.

We can now compare our oscillation result with the standard formula Eq.~\eqref{standardoscprof}. The mixing angle is $\abs{U_a}^2$ since we have assumed that source and detector both couple to the same linear combination of $\phi$. The imaginary part of the exponential is the same as Eq.~\eqref{standardoscprof} with $\xi=1/2$. The phase factor is the same as that obtained using $p_a L$, where $p_a = \sqrt{\Delta_S^2-m_a^2}$. The damping term is interesting. It says that the detector sees the source at the earlier time $ \tau - L/v_a$, which depends on the neutrino species $a$. Thus the different neutrino species are emitted from the source at different times, in such a way that they arrive at the detector at the same time. This is very different from the usual discussion, in which the source emits a neutrino wavepacket at $t=0$, which then propagates and spreads so that the different components arrive at the detector at different times. The consequences of Eq.~(\ref{oscprob1}) are  discussed further in Sec.~\ref{sec:conc}. The geometrical picture in Fig.~\ref{spacetimefig}  leads to the phase and damping factors in Eq.~(\ref{oscprob1}). We derive this in the next section, after we consider a moving source.

In the above calculation, the exponential decay of the source  damps the oscillations for baselines beyond $L_\text{coh} \sim (\Delta_S)^2/(\Gamma_S(m_a^2 - m_b^2))$, where $m_a$ and $m_b$ are two different neutrino masses.   For example, for a pion source producing $ \Delta_S \sim 1$ GeV neutrinos, the ratio of the coherence length  $L_\text{coh}$ to the oscillation length $L_\text{osc}=2 E_\nu/(m_a^2-m_b^2)$ is $L_\text{coh}/L_\text{osc} \sim \Delta_S/\Gamma_S \sim 10^{18}$.

\section{Oscillations with a Moving Source}
\label{movingsource}

In neutrino oscillation experiments with pion beams, the source moves relativistically in the lab frame. Using the model in Section~\ref{stationarysource}, we calculate the transition amplitude  with a source which begins  at the origin at $t=0$, and moves with a velocity ${\bf v}$ in the lab frame.  The detector remains stationary.  The relevant terms in the interaction Hamiltonian in Eq.~\eqref{simpleV} are modified to:
\begin{equation}
V(t) = \sum_a g \bigg[   U_a ~e^{-i\Delta_{S} t/\gamma}~ \phi^{(-)}_a(t,{\bf v}t) ~\ket{S_G}\bra{S_{E}} +  \sum_{j} U_a^*~ e^{i\Delta_{D_j} t}~\phi^{(+)}_a(t,{\bf L}) ~ \ket{D_{E_j}}\bra{D_G}  \bigg] 
+\text{h.c.}.
\end{equation}
The dependence on ${\bf v}$ and $ \gamma \equiv ( 1 - {\bf v}^2)^{-{\frac{1}{2}}}$ is determined by special relativity and is derived from a microscopic quantum field theory in Appendix \ref{heavystates}. The analogous expression for the transition amplitude in Eq.~\eqref{simpleamp} becomes:
\begin{eqnarray}
\mathcal{A}_j = -g^2  \int_{\tau}^{\tau+T} dt_1 \int_0^{t_1} dt_2 \int \frac{d^3 {\bf p}}{(2\pi)^3} \frac{1}{2E_{\bf p}}~e^{it_1(\Delta_D - E_{\bf p}) - it_2(\Delta_S/\gamma - E_{\bf p} + {\bf p \cdot v}) + i{\bf p\cdot L}} .
\end{eqnarray}
One can evaluate these integrals following the same procedure outlined in Section~\ref{stationarysource}.  In order to illustrate a specific example, we take ${\bf L} = L\hat{z}$ and ${\bf v} = v\hat{z}$.  The calculation is a bit tedious in three spatial dimensions due to the interplay between the momentum and angular integrals.  In order to illustrate the essential behavior of the oscillation probability, we do the calculation in one spatial dimension:
\begin{align}
\label{30}
\mathcal{A}_j = -g^2  \int_{\tau}^{\tau+T} dt_1 \int_0^{t_1} dt_2 \int_{-\infty}^\infty \frac{d{ p}}{2\pi} \frac{1}{2E_{ p}}~e^{it_1(\Delta_D - E_{ p}) - it_2(\Delta_S/\gamma - E_{ p} + { p  v}) + i{ p L}} .
\end{align}
Using the methods of Section~\ref{stationarysource}, the amplitude to excite detector $j$ inside the light-cone when $L > v \tau$ (so that the source does not hit the detector) is
\begin{equation}
\label{35} 
\mathcal{A}_j =  g^2  \sum_a |U_a|^2 \frac{ \gamma}{\sqrt{\Delta_S^2-m_a^2}}{e^{iL\sqrt{(E_{a})^2 - m_a^2} + i(\tau+T/2)(\Delta_D - E_{a})}} \left(\frac{\sin\left((\Delta_D - E_{a})T/2\right)}{\Delta_D - E_{a}}\right) .
\end{equation}
Here $E_{a} \equiv \gamma\left(\Delta_S + v\sqrt{\Delta_S^2 - m_a^2}\right)$ is the energy of the neutrino in the lab frame.  Eq.~\eqref{35}  differs from the one in three spatial dimensions by a factor of $\sqrt{\Delta_S^2 - m_a^2}$ in the denominator instead of $L^2$. There is no fall-off with $L$ in one dimension. Importantly, because the  energy of the neutrino now depends on its mass, our ideal model of a detector is capable, in principle, of distinguishing these different energies, and oscillations need not occur. For this to be possible, we need $\Gamma_S$ and $1/T$ to both be much smaller than the neutrino energy difference. Real-world detectors are far from ideal, so if there are two neutrinos with energy difference of order $(m_2^2 - m_1^2)v\gamma/\Delta_S$, and this difference is small enough to excite the same detector particle, then oscillations can occur. The oscillation probability can be estimated as in the previous section, making the replacement $\Delta_S \rightarrow \Delta_S - i\Gamma_S/2$ expanding in $m_a^2$, using $\Gamma_S \ll \Delta_S$, and picking out the term analogous to Eq.~\eqref{oscprob1} using Eq.~(\ref{21}),
\begin{equation}
\label{movingoscprob}
\mathcal{P}_\text{osc} \simeq \left| \sum_a |U_a|^2 e^{\frac{-im_a^2\gamma(L-v\tau)}{2\Delta_S} -\frac{\Gamma_S}{2\gamma}\left( \frac{\tau - L}{1-v} - \frac{m_a^2\gamma^2(L-v\tau)}{2\Delta_S^2} \right)} \right|^2 \,,
\end{equation}
where we have assumed that the neutrino energy difference is small compared with $\Gamma_S$, i.e.\ the experimental setup cannot resolve the energy difference. Equation~\eqref{movingoscprob} reduces to Eq.~\eqref{standardoscprof} with $\xi=1/2$ in the limit that $\tau \approx L$ and $\Gamma_S/\Delta_S$ is small, with $\Delta_S$ replaced by the relativistic Doppler shifted value
\begin{align}
\Delta_S \sqrt{\frac{1+v}{1-v}}\,.
\end{align}
We can set $\tau \approx L$ because if $\tau \gg L$, the damping factor makes the amplitude negligible.

The exponential decay factor in Eq.~\eqref{movingoscprob} has a simple geometrical interpretation, as in Fig.~\ref{spacetimefig}.
$\Gamma_S/\gamma$ is the inverse lifetime of the source in the lab frame.  Using straight line paths for the source and neutrino with velocity $v$ and $v_a$ respectively, the emission time $t_a$ (i.e. the time component of $x_1$ or $x_2$ in the figure) is
\begin{align}
\label{39}
t_a = \frac{v_a \tau - L}{v_a - v}\,.
\end{align}
The neutrino velocity is
\begin{align}
\label{40}
v_a &= \frac{\sqrt{E_a^2 - m_a^2}}{E_a}, & E_a = \gamma\left( \Delta_S + v\sqrt{\Delta_S^2 - m_a^2}\right)\,.
\end{align}
Substituting Eq.~(\ref{40}) in Eq.~(\ref{39}) and expanding in $m_a^2$ gives the coefficient of $\Gamma_S/(2\gamma)$ in Eq.~(\ref{movingoscprob}).

The neutrino propagation phase from Fig.~\ref{spacetimefig} is
\begin{align}
\label{41}
-i p_a \cdot (x_D-x_a) &=-i E_a (\tau-t_a)+ i p_a (L-v t_a)\,.
\end{align}
Expanding this in $m_a$ gives
\begin{align}
\label{42}
-i \frac{m_a^2}{\Delta_S}\gamma (L - v \tau)\,,
\end{align}
which is {\it twice} the phase in Eq.~\eqref{movingoscprob}.  However, there is another contribution to the neutrino phase from the phase of the source
\begin{align}
\label{43}
-ip_S \cdot x_S  &=  - i\Delta_S (\gamma t_a - \gamma v^2 t_a)
= -i \gamma \Delta_S(1+v) (\tau - L) + i \frac{m_a^2}{2 \Delta_S} \gamma (L-v \tau ).
\end{align}
The first term is independent of neutrino type. The second term cancels half the contribution in Eq.~\eqref{42}, to give the standard oscillation result in Eq.~\eqref{standardoscprof} with $\xi=1/2$. Eq.~\eqref{42} is the $\xi=1$ calculation outlined in Appendix~\ref{puzzle}, and is the correct neutrino propagation phase. However, the total neutrino phase at the detector also involves the phase of the source, which converts $\xi=1$ to $\xi=1/2$.  The usual derivation in the literature~\cite{Patrignani:2016xqp} of the $\xi=1/2$ value outlined in  Appendix~\ref{puzzle} ignores the phase of the source, and assumes neutrinos are emitted at the same time.

\section{Discussion and Conclusions}\label{sec:conc}

The source and detector in neutrino oscillation experiments can be systematically treated using heavy particle effective field theory.   In the heavy-particle limit, the source and the detector follow straight paths through spacetime, along which neutrinos can be created or absorbed.
Corrections to this treatment of the source can systematically be applied using the $1/M$ expansion, and in general can lead to a damping of the oscillations.  However, the baseline $L$  required to observe such damping from these corrections is many orders of magnitude larger than the diameter of Earth, so this is a negligible effect for contemporary neutrino oscillation experiments.

Oscillation experiments are typically performed with neutrinos that propagate a large spatial distance $L$ such that $E_\nu \gg 1/L$.  The neutrino propagator is then dominated by its pole. This allows us to illustrate the salient features of neutrino oscillation, treating the neutrino as a scalar and the source and detector system as well-localized, two-level systems, which undergo no recoil upon emitting or absorbing a neutrino.  Including the source, neutrino, and detector simultaneously, and performing standard time-dependent perturbation theory, we calculate the oscillation behavior both when the source particle is  stationary in the  lab frame and when it moves.    In the limit that $E_\nu \gg m_\nu$, $E_\nu \gg 1/L$, and the source does not traverse a large distance (compared to the oscillation length scale) before it decays, we recover the standard oscillation formula used by oscillation experiments.  Our treatment of neutrino oscillations can also be applied to mesonic oscillations.

Our calculations show that the detector sees the source at different earlier times, due to the different masses of the neutrinos.   This differs from statements frequently found in the literature, e.g., Refs.~\cite{Nussinov:1976uw, Kayser:1981ye, Giunti:1991sx, Giunti:2002xg, Giunti:1993se, Grimus:1996av, Kiers:1995zj, Giunti:1997wq, Stockinger:2000sk, Beuthe:2001rc, Beuthe:2002ej, Giunti:2003ax, Bernardini:2004sw, Bernardini:2006ak, Akhmedov:2008jn,  Cohen:2008qb, Akhmedov:2009rb, Naumov:2010um, Akhmedov:2010ms, Akhmedov:2012uu, Jones:2014sfa, An:2016pvi}, where it is said that all neutrino mass eigenstates are produced at the moment the source decays, and the different mass eigenstates propagate with different velocities in the lab frame.  Importantly, this is said to lead to an inescapable source of damping of neutrino oscillations at large values of $L$, since the neutrino wave packets associated with different mass eigenstates separate between production and detection due to their different velocities and may no longer coherently interact with the detector. This is different from our conclusion, that the neutrinos are emitted at different times, such that they arrive at the detector simultaneously, as in Fig.~\ref{spacetimefig}.   To illustrate the differences between our results and those in the literature:~if the source is stationary, maintaining a state unchanged in time, and the particles produced along with the neutrino when the source decays are unmeasured, the common expectation is that oscillations would not persist in the limit $L\rightarrow \infty$.  However, we find no damping of the oscillations occur in this scenario.  In the case of $\Gamma_S \not =0$, the conventional wavepacket picture says that the lighter neutrino wavepacket arrives first at the detector, followed by the heavier neutrino wavepacket. Our calculation shows that the lighter neutrino amplitude is suppressed by $e^{-\Gamma_S \Delta t/2}$, where $\Delta t$ is the difference in transit times, because it is emitted later. Thus the detector sees a ``brighter'' heavy neutrino emission,  simultaneously with a ``fainter'' light neutrino.

One might think that the oscillation problem has a symmetry given by reversing the direction of time and swapping source and detector.  However, in Fig.~\ref{spacetimefig}, one fixes the {\it initial} state of the quantum system, and evolves it in time. In the time-reversed process, this means the {\it final} state rather than the initial state is fixed, which is not the case in  experiments. This is an intrinsic asymmetry in the boundary conditions that breaks the time reversal symmetry.

A more realistic treatment of the physics underlying neutrino oscillation experiments can be given using the framework presented here.  For example, the neutrino is a fermion, the source has a wave function (though much smaller than any oscillation length), neutrinos are produced and absorbed via the weak interactions, etc.  Effects such as these lead to different production and absorption probabilities, but they retain the same oscillation behavior.  There are additional physical effects that are present in reactor and pion-beam experiments that can lead to damping of oscillations.   These include any treatment of the source and detector that differs from free evolution.  In the calculations presented in Sections~\ref{stationarysource} and~\ref{movingsource}, we included the lifetime of the source as a means by which the source does not undergo pure free evolution, but gets exponentially damped in time.  Other effects can include small movements of the nucleus in reactor experiments and collisions between the pion and air molecules in accelerator experiments, both of which would lead to damping of the neutrino oscillations.  In optics, these are called broadening mechanisms, and have been studied at length (see, for example, Ref.~\cite{Mandel:1995seg,Loudon:2000}).   These effects are expected to be negligible for current terrestrial neutrino oscillation experiments.  Development of experimental setups where such damping of oscillations is observable would be a valuable verification of the understanding of the quantum mechanical behavior of neutrino oscillations and conceivably could be used to measure previously-unmeasured properties
of neutrinos, such as the individual masses of the neutrinos. 

\begin{acknowledgments}
We would like to thank P.~Stoffer for comments on the manuscript. This work is supported in part by DOE grant \#DE-SC0009919. 
\end{acknowledgments}

\appendix

\section{The Standard Neutrino Oscillation Derivation}
\label{puzzle}

The following is in the spirit of the calculation of the behavior of neutrino oscillation presented in the PDG~\cite{Patrignani:2016xqp}.  Consider a single neutrino flavor eigenstate $\ket{\nu_\ell}$, which is a superposition of mass eigenstates $\ket{\nu_a}$:
\begin{align}
| \nu_\ell \rangle &= \sum_a U^*_a |\nu_a\rangle, & \sum_a |U_a|^2 &= 1.  
\end{align}
On their journey through spacetime, the mass eigenstates all propagate a time $T$ and distance $L$:
\begin{equation}
| \nu_\ell (T,L) \rangle =  \sum_a U^*_a e^{-i(E_aT - p_aL)} | \nu_a \rangle \, ,
\end{equation}
and the amplitude for the neutrino to transition from flavor $\ell$ to $\ell'$ over time $T$ and distance $L$ is:
\begin{equation}
\mathcal{A}(T,L) = \langle \nu_{\ell'}(T,L) | \nu_{\ell} \rangle \, .
\end{equation}
The oscillation probability is $\mathcal{P}(T,L) = |\mathcal{A}(T,L)|^2$.  Because the neutrinos are ultra-relativistic, and if, for simplicity, the neutrino is produced in a $1\rightarrow 2$ process, $S \to \nu + X$, the following approximations can be made for the momentum $p_a$ and energy $E_a$ for each mass eigenstate:
\begin{align}
p_0 &= \frac{m_S^2-m_X^2}{2m_S}, \\
\label{pa}
p_a &\simeq p_0 - \left(1 + \frac{m_X^2}{m_S^2}\right) \frac{m_a^2}{4p_0} + \mathcal{O}\left(\frac{m_a^4}{p_0^3} \right),  \\
\label{Ea}
E_a &= \sqrt{ p_a^2 + m_a^2 }~ \simeq ~p_0 + \left(1-\frac{m_X^2}{m_S^2} \right) \frac{m_a^2}{4p_0} + \mathcal{O}\left( \frac{m_a^4}{p_0^3} \right).
\end{align}
The usual derivation of the neutrino oscillation expression uses the neutrino phase
\begin{align}
\label{A7}
-i \left( E_a T - p_a L\right) &\simeq - i \frac{m_a^2}{2 p_0}L, \qquad \hbox{using $T=L$}\,,
\end{align}
where all neutrino species travel the same distance and the same time $T=L$, and we have expanded to quadratic order in $m_a$. This equation is the one utilized by experiments when extracting measured values of the neutrino oscillation parameters.  The derivation also holds for $1 \to \text{many}$ decays, where $X$ is a multiparticle state with invariant mass $m_X$.  Note that one does not have to assume ``equal energy'' nor ``equal momentum'' among neutrino mass eigenstates, which is a parlance sometimes found in the literature.  

There are, however, two issues with this standard derivation.  First, the justification why one must choose that $T = L$ is not a result of a calculation.  One could instead use for $T_a$ the neutrino time of flight in the lab frame:
\begin{equation}
T_a = \frac{L}{v_a} = \frac{E_a}{p_a}L \,.
\end{equation}
This correction in the time of flight can be very small, but including the corrections in Eqs.~\eqref{pa} and \eqref{Ea} lead to an oscillation phase that differs from Eq.~\eqref{A7} in oscillation frequency by a factor of 2: 
\begin{align}
\label{A8}
-i \left( E_a T_a - p_a L\right) \simeq - i \frac{m_a^2}{p_0}L\,,
\end{align}
It is not clear which oscillation phase is correct, Eq.~\eqref{A7} or Eq.~\eqref{A8},  since the only choice was to what order a Taylor expansion is carried out.  
The second issue is that the oscillation probability $\mathcal{P}(T,L)$ depends on the neutrino transit time $T$.  This time is not a measured observable, since there is no information extracted from standard neutrino oscillation experiments regarding when the source particle decays.  Therefore, one cannot assign a value to $T$.  Instead, one must sum over the amplitudes for all neutrino transit times.  If one does this, as explicitly calculated using a simple system in Sections~\ref{stationarysource} and~\ref{movingsource}, and using a more general system in Appendix~\ref{fullqftrecoil}, the oscillation phase is unambiguously calculated to be the one in Eq.~\eqref{A7}, and the oscillation amplitude  is dominated by terms associated with the source decaying at different times, such that neutrinos of different mass interfere at the detector at the same time, as illustrated with the cartoon in Fig.~\ref{spacetimefig}.

\section{Effective Interactions for a Heavy Source Particle}
\label{heavystates}

\subsection{Initial state of the source}
\label{intialstateofsource}

In neutrino oscillation experiments, the source particle that produces  the neutrino must be  relatively well-localized in both position and momentum~\cite{Cohen:2008qb}. It is important that the spread in momentum is not zero,  so that the source may be localized to a distance much less than the oscillation length: $L_\text{osc} \gg \Delta x .$ As an example, for pion beam experiments with energy of order a GeV:
\begin{align}
L_\text{osc}  = \frac{ 2 E }{\Delta m^2 } \sim \mathcal{O}(100 \text{ km})\,,
\end{align}
which is much larger than the typical size of the source.

The initial state of the position degree of freedom of the source  can be approximated as a Gaussian wave function with velocity ${\bf v}$ centered at ${\bf x}_0$:
\begin{align}
\label{wavepacketrevenge}
\ket{\psi(t=0)} &= \int d^d{\bf x}~\psi(t=0,{\bf x})~ \ket{{\bf x}}_\text{NR}  \nn 
&=
 \int d^d{\bf x} ~\left( \frac{1}{\sqrt{2\pi} \Delta x} \right)^{d/2} e^{-\frac{({\bf x} - {\bf x}_0)^2}{4\Delta x^2} + i\gamma M {\bf v \cdot (x} - {\bf x}_0)} ~\ket{\bf {x}}_\text{NR} \nn  
&= \int \frac{d^d {\bf p}}{(2\pi)^d} \left(2\sqrt{2\pi}\Delta x \right)^{d/2} e^{-({\bf p} - \gamma M {\bf v})^2 \Delta x^2 - i{\bf p \cdot x}_0} ~\ket{{\bf p}}_\text{NR} \, \nn 
& \equiv \int \frac{d^d {\bf p}}{(2\pi)^d}~ \widetilde{\psi}({\bf p}) ~\ket{{\bf p}}_\text{NR}  ,
\end{align}
where $d$ is the number of spatial dimensions, $\gamma$ is the Lorentz factor $\gamma \equiv (1-{\bf v}^2)^{-1/2}$, and the subscripts NR indicate non-relativistic state normalization.\footnote{The single particle states here have nonrelativistic normalization:
$$
\ket{{\bf p}}_\text{NR} \equiv a_{\bf p}^\dagger \ket{0} = \int d^d{\bf x} ~e^{i{\bf p\cdot x}} ~\ket{{\bf x}}_\text{NR}, \hspace{0.25in} [a_{\bf k}, a^\dagger_{\bf p}] = (2\pi)^d \delta^d({\bf k} - {\bf p}), \hspace{0.25in} \braket{{\bf k} | {\bf p}}_\text{NR} = (2\pi)^d \delta^d({\bf k} - {\bf p}). \nonumber
$$
}
This state is localized at $\mathbf{x}_0$ with velocity $\mathbf{v}$, and we can make the uncertainties $\Delta x$ and $\Delta v$ arbitrarily small by taking $\Delta x \to 0$ and $M \to \infty$ with $M \Delta x \to \infty$.

Under free time evolution, the state evolves to 
\begin{eqnarray}
\ket{\psi(t)} &=& e^{-it\sqrt{\hat{\bf p}^2 + M^2} } \ket{\psi(t=0)} = \int d^d{\bf x}~ \psi(t, {\bf x}) ~ \ket{{\bf x}}_\text{NR} ,
\end{eqnarray}
where now 
\begin{eqnarray}
\psi(t,{\bf x}) &=& \int \frac{d^d{\bf p}}{(2\pi)^d} ~ \widetilde{\psi}({\bf p})~e^{-it\sqrt{{\bf p}^2 + M^2} + i {\bf p \cdot x}}\,,\nn 
&=& \left(2\sqrt{2\pi}\Delta x \right)^{d/2} \int \frac{d^d{\bf p}}{(2\pi)^d} ~e^{-({\bf p} - \gamma M {\bf v})^2 \Delta x^2 -it\sqrt{{\bf p}^2 + M^2} + i {\bf p \cdot( x-x}_0)}\,.
\end{eqnarray}
We can expand the argument of the exponent about ${\bf p } = \gamma M {\bf v} + {\bf k}$:
\begin{align}
& -({\bf p} - \gamma M {\bf v})^2 \Delta x^2 -it\sqrt{{\bf p}^2 + M^2} + i {\bf p \cdot( x-x}_0)  \nonumber \\
&= - i \gamma M (t-{\bf v}\cdot({\bf x} - {\bf x}_0)) +i ({\bf x} - {\bf x}_0 - {\bf v}t) \cdot {\bf k} - \left( \Delta x^2 + \frac{it}{2M\gamma^3} \right) {\bf k}^2 + ...
\end{align}
where we neglect higher-order terms in ${\bf k}$, since their effects are suppressed by higher powers of $1/(M\Delta x)$. Performing the Gaussian integral over ${\bf k}$,
\begin{align}
\psi(t,{\bf x}) &= \left( \frac{\Delta x}{\sqrt{2\pi} \alpha(t) }\right)^{d/2} e^{ -\frac{({\bf x} - {\bf x}_0 - {\bf v}t)^2}{4\alpha(t)} - i \gamma M (t-{\bf v}\cdot({\bf x} - {\bf x}_0))} , & \alpha(t) &\equiv \Delta x^2 + \frac{it}{2 M\gamma^3} .
\end{align}
In the regime where $t/(M\Delta x^2) \ll 1$, the center of the wave function moves with velocity ${\bf v}$ and the wave function does not spread with time.  Neglecting this term is justified by the fact that $e^{-\Delta x^2 {\bf k}^2}$ suppresses the large-$k$ integration region, so that $k \sim 1/\Delta x$, and the term in question scales like
\begin{equation}
\frac{it}{2M\gamma^3}k^2 \sim \frac{it}{2M\gamma^3\Delta x^2} .
\end{equation}
This is negligible if one takes $M\Delta x^2 \rightarrow \infty$.  For experimental sources of neutrinos, e.g., charged pions in accelerator experiments and radioactive nuclei in reactor experiments, the momentum uncertainty $\Delta x^{-1}$ is very small compared to the source's Compton wavelength, as discussed in Section~\ref{heavysources}, and we can ignore the free-particle diffusion of the state $\ket{\psi(0)}$.  If so, the wave function simplifies to
\begin{align}
\label{b8}
\psi(t,{\bf x}) = \left( \frac{1}{\sqrt{2\pi}\Delta x  }\right)^{d/2} e^{ -\frac{({\bf x} - {\bf x}_0 - {\bf v}t)^2}{4\Delta x^2} - i \gamma M (t-{\bf v}\cdot({\bf x} - {\bf x}_0))} ,
\end{align}
which, up to a phase, is the same wave function at $t=0$, except displaced by an amount ${\bf v}t$.  It will be useful to note that $- i \gamma M (t-{\bf v}\cdot({\bf x} - {\bf x}_0)) = - i Mt/\gamma + i\gamma M {\bf v} \cdot({\bf x} - {\bf x}_0 -{\bf v}t)$.   If we simultaneously take $\Delta x \rightarrow 0$, $M\Delta x \rightarrow \infty$, and $M\Delta x^2 \rightarrow \infty$, then $|\psi(t,{\bf x})|^2$ can be approximated as a $\delta$-function:
\begin{equation}
\lim_{M\rightarrow \infty }|\psi(t,{\bf x})|^2 \sim \delta^d({\bf x} - {\bf x}_0 - {\bf v}t) .
\end{equation}

\subsection{Heavy-to-heavy transitions}

Consider the region of parameter space where the source decays to a neutrino and another heavy state, at zero recoil.  In this case, we may label the heavy states by an internal label $\alpha = G, E$, indicating the level of excitation.  The energy difference between these two states is the quantity $\Delta_S$ appearing in Sections~\ref{stationarysource} and~\ref{movingsource}, i.e., the energy imparted to the daughter neutrino.  The regime under discussion here is $\Delta_S \ll M_S$, where $M_S$ is the mass of the $E$ state of the source.\footnote{This regime was explored in Ref.~\cite{Shifman:1987rj} in the context of $b\rightarrow c$ transitions.  }   Standard arguments show how, in this regime, the position degree of freedom of the source particle, initially in a state of the form in Eq.~\eqref{wavepacketrevenge}, can be treated as a fixed straight-line trajectory, up to corrections $1/M$.  We review this logic briefly, following Ref.~\cite{Peskin:1983up}.  Consider a scalar theory with fields $\phi$ and $\Phi_\alpha$, where $\alpha = G, E$.  The action is
\begin{equation}
S[\phi] = \int d^{d+1}x ~\frac{1}{2} \left( \partial_\mu \Phi_\alpha \partial^\mu \Phi_\alpha - M_\alpha^2 \Phi_\alpha^2 - g~\phi(x)~ \Phi_\alpha X_{\alpha\beta} \Phi_\beta \right) ,
\end{equation}
where $X_{\alpha\beta} = \left(\begin{array}{cc} 0 & 1 \\ 1 & 0  \end{array} \right)$,  and $g$ is a coupling constant.  For the moment, we treat $\phi(x)$ as a fixed background field that acts as a spacetime-dependent mass for the dynamical field $\Phi$.  The two-point Green's function, $G_{\alpha\beta}(x,y) \equiv \bra{\Omega} \mathcal{T} \Phi_\alpha(x) \Phi_\beta(y) \ket{\Omega}$, satisfies the Schwinger-Dyson equation:
\begin{equation}
-i \delta^{d+1}(x-y) \delta_{\alpha\gamma} = \left\{ (\partial^2- M_\alpha^2)\delta_{\alpha\beta} - gX_{\alpha\beta}\phi(x) \right\} G_{\beta \gamma} (x,y) .
\end{equation}
The solution to this equation has the path integral representation
\begin{equation}
G(x,y) = \int_0^\infty dT \int_{x(0)=x}^{x(T) = y} [\mathcal{D}x] ~e^{-\frac{i}{2} \int_0^T d\tau \left( \dot{x}^\mu \dot{x}_\mu + M^2 + gX\phi(x) \right) }. 
\end{equation}
In a non-relativistic situation, where the spacetime points $x$ and $y$ are separated by a timelike distance large compared to $1/M$, the integral is dominated by the stationary-phase configuration, which is a straight-line path from $x$ to $y$.  The exponent in $e^{-iS_1[x]}$ evaluated on the stationary path $\underline{x}$ (treating $g$ as a perturbation, and performing the integral over $T$ by the stationary phase as well) is
\begin{equation}
S_1[\underline{x}] = \frac{M}{\gamma} t_f + gX\int_0^{t_f/M} d\tau ~ \phi(\underline{x}(\tau)) , 
\end{equation}
where $t_f \equiv\sqrt{(x-y)^2}$,  $\gamma = (1-{\bf v}^2)^{-1/2}$, and ${\bf v } \equiv \frac{{\bf x} - {\bf y}}{x^0 - y^0}$.  In a sector where there is a single $\Phi$ particle, we can simply add this term to the action for the $\phi$ field.  The first term is just the phase accumulated by the rest energy of the $\Phi$ particle.  The second term can be rewritten as
\begin{equation}
\label{eq:effective-V}
- \int dt ~V(t) = gX\int_0^{t_f/M} d\tau ~ \phi(\underline{x}(\tau)) = g \left( \ket{S_G} \bra{S_E} + h.c. \right)\int dt ~d^d{\bf x} ~\delta^d({\bf x} - {\bf v}t)~ \phi(t, {\bf x}), 
\end{equation}
which is the effective interaction we have used in Sections~\ref{stationarysource} and~\ref{movingsource}.

\subsection{Heavy-particle states including recoil}
\label{fullqftrecoil}

If, as a result of the interaction that produces the neutrino, the source particle annihilates into a set of lighter particles, we can no longer treat its trajectory as a straight line throughout the process. Rather, it should be a straight line which ends at the interaction event.  This is similar to the situation in HQET of $b \to u$ decays, mediated by the current $\overline u\, \Gamma\, b_v$~\cite{Manohar:2000dt}. $b_v$ annhilates the $b$ quark moving with velocity $v$, and $\overline u$ creates the light quarks in the final hadron. Here, we show that  the essential form of the amplitude  (in particular the phase which leads to the neutrino oscillations) is unchanged as a result of this generalization.

We infer the interaction vertex from a local, translation-invariant interaction vertex between the neutrino\footnote{See footnote~\ref{foot:fermions} for an explanation of why oscillation behavior of a fermion is the same as that of a scalar.} field $\phi(x)$, and an operator $\osx$ that annihilates the source state and creates the $X$ state, which is the set of particles produced in conjunction with the neutrino:
\begin{equation}
V(t) = g_S \int d^d{\bf x} ~\phi^{(-)}(t, {\bf x})\osx(t, {\bf x})  + h.c. + g_D \ket{D_E} \bra{D_G} + h.c. \,.
\end{equation}
We have kept the detector infinitely heavy, as in the main text, and used the rotating wave approximation for $\phi$, which is valid for calculations within the light-cone for long baselines. For example, for neutrino production by $\pi \to \mu \overline \nu_\mu$ decay due to the
weak Hamiltonian
\begin{align}
H_W &= \frac{4 G_F}{\sqrt 2} (\overline \nu_\mu \gamma^\mu P_L \mu)(\overline d \gamma_\mu P_L u)\,,
\end{align}
$\osx$ is
\begin{align}
\label{b17}
g_S \osx &= \frac{4 G_F}{\sqrt 2}\sum_{p_\mu,s_\mu, p_\pi}  \ket{\mu,p_\mu,s_\mu}\braket{\mu,p_\mu,s_\mu | (\gamma^\mu P_L \mu)_\alpha (\overline d \gamma_\mu P_L u)|\pi,p_\pi}\bra{\pi,p_\pi}\,,
\end{align}
given by projecting onto all $\pi$ initial states and all $\mu$ final states. The $S \to X$ transition is given by the form factor
$F_S$ defined as
\begin{equation}
F_S(p_S, p_X) \equiv \bra{X} \osx (0, 0) \ket{{\bf p}_S}\,,
\end{equation}
so that
\begin{align}
\label{b19}
\osx &= \sum_{X,\mathbf{p}_S} F_S(p_S,p_X) \ket{X,\mathbf{p}_X} \bra{\mathbf{p}_S}\,.
\end{align}
The spin labels on $X$ are included in $p_X$. $F_S$ has an additional spinor label $\alpha$ for a fermionic neutrino.
Exponential decay of the source can be introduced by the substitution $E_{{\bf p}_S} \rightarrow E_{{\bf p}_S}  - i \Gamma_S/2$, where $1/\Gamma_S$ is the lifetime.

The leading-order amplitude for neutrino exchange between the source  (in the initial state $\ket{\psi(t=0)}$ with momentum space wave function  $\widetilde{\psi}({\bf p}_S)$) and the detector, with an $X$ state of total momentum $p_X$ in the final state, is then
\begin{align}
\mathcal{A}(p_X) &= (-i)^2 g_S g_D \int_{\tau}^{\tau+T} dt_1 \int_0^{t_1} dt_2 ~ \bra{D_E; 0_\phi; X} 
V(t_1)V(t_2)  \ket{D_G ; 0_\phi; \psi(t=0)} \nonumber \\
&= - g_S g_D \int_{\tau}^{\tau+T} dt_1 \int_0^{t_1} dt_2 \int d^d{\bf x} \int \frac{d^d{\bf p}_S}{(2\pi)^d} ~ F_S(p_X, p_S) ~e^{i \Delta_D t_1 + i(E_X - E_{{\bf p}_S})t_2 - i({\bf p}_X - {\bf p}_S)\cdot {\bf x}} \nonumber \\
& \hspace{4cm} \times~ \bra{0} \phi^{(+)}(t_1,{\bf L}) \phi^{(-)}(t_2, {\bf x}) \ket{0} ~\widetilde{\psi}({\bf p}_S) .
\end{align}
Using the representation
\begin{equation}
\bra{0} \phi^{(+)}(t_1,{\bf L}) \phi^{(-)}(t_2, {\bf x}) \ket{0} = \int \frac{d^d{\bf p}}{(2\pi)^d} \frac{1}{2E_{\bf p}} e^{-iE_{\bf p}(t_1 - t_2) + i {\bf p \cdot (L-x)}} ,
\end{equation}
\begin{align}
\mathcal{A}(p_X) 
&= - g_S g_D \int_{\tau}^{\tau+T} dt_1 \int_0^{t_1} dt_2 \int d^d{\bf x} \int \frac{d^d{\bf p}_S}{(2\pi)^d}  \int \frac{d^d{\bf p}}{(2\pi)^d} \frac{1}{2E_{\bf p}}~\widetilde{\psi}({\bf p}_S) \nonumber \\
& \qquad \times F_S(p_X, p_S) ~e^{i ( \Delta_D - E_{\bf p}) t_1 + i(E_{\bf p} + E_X - E_{{\bf p}_S})t_2 - i({\bf p}+{\bf p}_X - {\bf p}_S)\cdot {\bf x}+ i \mathbf{p \cdot L}} \,.
\end{align}
The initial state of the source $\widetilde{\psi}$ is given in Section~\ref{intialstateofsource},  Eq.~(\ref{wavepacketrevenge}). Doing the $\mathbf{p}_S$ integral,
\begin{eqnarray}
\label{b23}
\int \frac{d^d{\bf p}_S}{(2\pi)^d} ~F_S(p_X, p_S)~\widetilde{\psi}({\bf p}_S) ~e^{-iE_{{\bf p}_S}t_2 + i{\bf p }_S \cdot {\bf x}}  = F_S(p_X,  M v^\mu)~ \psi(t_2, {\bf x}) + \mathcal{O}\left( \frac{1}{M \Delta x} \right).
\end{eqnarray}
Here, $v^\mu = (\gamma, \gamma{\bf v})$.  This evaluation followed the same logic as in Section~\ref{intialstateofsource}.  We assumed that $F_S$ is analytic in $p_S$, and varies on a typical hadronic scale $\Lambda_{\rm QCD}$, which is much bigger than $1/\Delta x$.  The source has momentum $M v^\mu$, with a small spread of order $1/\Delta x$. Since the form factor is almost constant over this momentum spread, $F_S$ can be factored out of the integral Eq.~(\ref{b23}). The remaining integral gives the time-evolved source, which we have seen in Section~\ref{intialstateofsource} moves with velocity $\mathbf{v}$, so that it is now centered at $\mathbf{x}_0 + \mathbf{v}t_2$, as in Eq.~(\ref{b8}). This gives Eq.~(\ref{b23}).
$\psi(t_2, {\bf x})$ is the time-evolved spatial wave function:
\begin{equation}
\psi(t_2, {\bf x}) = \left( \frac{1}{\sqrt{2\pi}\Delta x  }\right)^{d/2} e^{ -\frac{({\bf x} - {\bf x}_0 - {\bf v}t_2)^2}{4\Delta x^2} - i M t_2/\gamma + i\gamma M {\bf v}\cdot({\bf x} - {\bf x}_0 - {\bf v} t_2)} .
\end{equation}
 The amplitude is then
\begin{align}
\label{b25}
\mathcal{A}(p_X) 
&= -g_S g_D \int_{\tau}^{\tau+T} dt_1 \int_0^{t_1} dt_2 \int d^d{\bf x}   \int \frac{d^d{\bf p}}{(2\pi)^d} \frac{1}{2E_{\bf p}} \nonumber \\
&\qquad \times F_S(p_X,  M v^\mu)~ \psi(t_2, {\bf x})
 ~e^{i ( \Delta_D - E_{\bf p}) t_1 + i(E_{\bf p} + E_X)t_2 - i({\bf p}+{\bf p}_X)\cdot {\bf x}+ i \mathbf{p \cdot L}} \,.
\end{align}
This is essentially the result derived in the text for a Wilson line source, multiplied by the form factor $F_S$. 
The significant technical difference is the presence of the wavefunction $\psi(x)$.  
In the following we show that the integrals can still be done, 
and the conclusions are unchanged.  The impatient reader 
is invited to skip to the final answer,
in Eq.~\eqref{eq:finally}.

From now on, we set $\mathbf{x}_0=0$ in $\psi(t_2, {\bf x})$, so that the source starts at the origin.
Isolating the ${\bf x}$ dependence of the integral Eq.~(\ref{b25}):
\begin{eqnarray}
\label{eq:first-step-of-x-integral}
\int d^d{\bf x} ~ e^{  - \frac{({\bf x}  - {\bf v}t_2)^2}{4\Delta x^2}  -i({\bf p}_X + {\bf p} - \gamma M {\bf v})\cdot {\bf x} } &=& (2\sqrt{\pi} ~\Delta x)^d ~e^{-\Delta x^2({\bf p} + {\bf p}_X - \gamma M {\bf v})^2 - i {\bf v}t_2 \cdot({\bf p} + {\bf p}_X - \gamma M {\bf v}) } \nonumber 
\\
&=& (2\pi)^d ~\delta^d_{1/\Delta x}({\bf p} + {\bf p}_X - \gamma M {\bf v})
~e^{ - i {\bf v}t_2 \cdot({\bf p} + {\bf p}_X - \gamma M {\bf v}) },
\end{eqnarray}
where 
$ \delta_{\Delta p}(p) \equiv { 1 \over \sqrt{\pi} \Delta p } e^{ - p^2/\Delta p^2  } $ is an approximate delta function with width $\Delta p$.
Eq.~(\ref{b25}) becomes
\begin{align}
\label{b25a}
{\cal A}(p_X) &= - g_S g_D \int_{\tau}^{\tau+T} dt_1 \int_0^{t_1} dt_2 \int \frac{d^d{\bf p}}{(2\pi)^d} \frac{1}{2E_{\bf p}} (2\pi)^d ~\delta^d_{1/\Delta x}({\bf p} + {\bf p}_X - \gamma M {\bf v})   \left( \frac{1}{\sqrt{2\pi}\Delta x  }\right)^{d/2} \nonumber \\  
& F_S(p_X,  M v^\mu)~ 
e^{it_1(\Delta_D - E_{\bf p}) + i t_2(E_X + E_{\bf p}- M/\gamma -  {\bf p}_X \cdot {\bf v}  - {\bf p \cdot v})  + i {\bf p} \cdot {\bf L}}\,.
\end{align}
The exponential $\Delta_S/\gamma$ in Eq.~(\ref{30}) has been replaced by the $M/\gamma - (E_X - \mathbf{p}_X \cdot \mathbf{v})$, the Doppler shifted  energy of the source transition, and the integrand has the form factor $F_S(p_X,Mv^\mu)$ and the approximate momentum conserving $\delta$-function $\delta^d_{1/\Delta x}({\bf p} + {\bf p}_X - \gamma M {\bf v})$.

For simplicity, we now set ${\bf v} = 0$,  ${\bf x}_0 = 0$, and $\Gamma_S = 0$, and proceed to study 
\begin{eqnarray}
\mathcal{A}(p_X) &=& -g_S g_D (2^{3/2} \sqrt{\pi} \Delta x)^{3/2} F_S(p_X, Mv^\mu)  \int_\tau^{\tau+T} dt_1 \int_0^{t_1} dt_2 \int \frac{d^3{\bf p}}{(2\pi)^3} \frac{1}{2E_{\bf p}} e^{-\Delta x^2({\bf p} + {\bf p}_X)^2}  \nonumber \\
&& \hspace{1in} \times e^{it_1(\Delta_D - E_{\bf p} ) + it_2(E_X + E_{\bf p} - M ) + i {\bf p \cdot L}}~.
\end{eqnarray}
Doing the $t_2$ integral, and letting ${\bf L} = L \hat{z}$:
\begin{eqnarray}
\mathcal{A}(p_X) &=& -ig_S g_D (2^{3/2} \sqrt{\pi} \Delta x)^{3/2}  F_S(p_X, Mv^\mu)  \int_\tau^{\tau+T} dt_1 ~e^{it_1\Delta_D} \int \frac{d^3{\bf p}}{(2\pi)^3} \frac{1}{2E_{\bf p}} e^{-\Delta x^2({\bf p} + {\bf p}_X)^2}  \nonumber \\
&& \hspace{1in} \times ~e^{ i {\bf p \cdot L}} \left[\frac{e^{-it_1 E_{\bf p} } - e^{-i(M - E_X)t_1}}{E_{\bf p} + E_X - M} \right] \\
&=& -ig_S g_D (2^{3/2} \sqrt{\pi} \Delta x)^{3/2}  F_S(p_X, Mv^\mu)  \int_\tau^{\tau+T} dt_1 ~e^{it_1\Delta_D} \int_0^\infty \frac{p^2dp}{(2\pi)^3} \frac{1}{2E_{ p}} \left[\frac{e^{-it_1 E_{ p} } - e^{-i(M - E_X)t_1}}{E_{ p} + E_X - M} \right]   \nonumber \\
&& \hspace{1in} \times \int_{-1}^1 dx \int_0^{2\pi}d\phi~e^{ i pLx-\Delta x^2\left[p^2 + {\bf p}_X^2 + 2p(p_X^x\sqrt{1-x^2} \cos\phi + p_X^y\sqrt{1-x^2} \sin\phi + p_X^z x)\right]} 
 \end{eqnarray}
%


\footnotesize
\noindent
{\bf Aside \#1}. It will be useful to estimate the value of the following integral, to leading order in $1/a$, $1/b$, $1/c$, and where $a \gg b,c$:
\begin{equation}
I = \int_{-1}^1 dx ~ e^{(ia+c)x + b\sqrt{1-x^2}}
\end{equation}
The integrand is highly peaked at $x=1$ if $c>0$.  Expanding the argument of the exponent around $x=1$, where $\rho^2 \equiv 1-x$:
\begin{eqnarray}
I &\sim& -2 e^{ia+c} \int_{\infty}^0 \rho~ d\rho ~ e^{\sqrt{2}b\rho-(ia+c)\rho^2 } \\
&=& -2 e^{ia+c} \left[ -\frac{1}{2(ia+c)} - \sqrt{\frac{\pi}{2}} \frac{be^{\frac{b^2}{2(ia+c)}}}{2(ia+c)^{3/2}}\left(1 + \text{Erf}\left( \frac{b}{\sqrt{2(ia+c)}} \right) \right) \right] \\
&=& \frac{e^{ia + c}}{ia} + \mathcal{O}\left( \frac{c}{a}, \frac{b}{a^{3/2}} \right)
\end{eqnarray}
If instead $c<0$, then the integrand is highly peaked around $x = -1$.   Expanding the argument of the exponent around $x=-1$, where $\rho^2 \equiv -1 + x$, 
\begin{eqnarray}
I &\sim& 2 e^{-ia-c} \int_{\infty}^0 \rho~ d\rho ~ e^{\sqrt{2}b\rho+(ia+c)\rho^2 } \\
&=& -\frac{e^{-ia -c}}{ia} + \mathcal{O}\left( \frac{c}{a}, \frac{b}{a^{3/2}} \right)
\end{eqnarray}
%


\normalsize

\noindent
Using the approximations in Aside \#1, one can do the angular integrals, in the limit that $\Delta x/L \ll 1$:
\begin{eqnarray}
\mathcal{A}(p_X) &=& -ig_S g_D (2^{3/2} \sqrt{\pi} \Delta x)^{3/2}  F_S(p_X, Mv^\mu)  \int_\tau^{\tau+T} dt_1 ~e^{it_1\Delta_D} \int_0^\infty \frac{p^2dp}{(2\pi)^3} \frac{1}{2E_{ p}} \left[\frac{e^{-it_1 E_{ p} } - e^{-i(M - E_X)t_1}}{E_{ p} + E_X - M} \right]   \nonumber \\
&& \hspace{1in} \times \left( \frac{2\pi}{ipL} \right) \times \left\{ \begin{array}{ll} e^{ipL - \Delta x^2(p \hat{z} + {\bf p}_X)^2}, & \text{if } p_X^z < 0  \\ -e^{-ipL - \Delta x^2(p \hat{z} - {\bf p}_X)^2},  & \text{if } p_X^z > 0 \end{array} \right.~~.
 \end{eqnarray}
 The point is that 
the angular integral over the direction of ${\bf p}$,
in the limit $ L \gg \Delta x^2 p$,
is dominated by the saddle point at 
$ {\bf p} = \hat L p + {\cal O}\left({ \Delta x^2 p \over L} \right)$ --  the neutrino comes from the source, up 
to corrections from the uncertainty in the position of the source.

Changing integration variables from $p \rightarrow E_p$:
\begin{eqnarray}
\mathcal{A}(p_X) &=& -ig_S g_D (2^{3/2} \sqrt{\pi} \Delta x)^{3/2}  F_S(p_X, Mv^\mu)  \int_\tau^{\tau+T} dt_1 ~e^{it_1\Delta_D} \int_m^\infty \frac{dE_p}{2(2\pi)^3} \left[\frac{e^{-it_1 E_{ p} } - e^{-i(M - E_X)t_1}}{E_{ p} + E_X - M} \right]   \nonumber \\
&& \hspace{1in} \times \left( \frac{2\pi}{iL} \right) \times \left\{ \begin{array}{ll} e^{iL\sqrt{E_p^2 - m^2} - \Delta x^2(\sqrt{E_p^2 - m^2} \hat{z} + {\bf p}_X)^2}, & \text{if } p_X^z < 0  \\ -e^{-iL\sqrt{E_p^2 - m^2} - \Delta x^2(\sqrt{E_p^2 - m^2} \hat{z} - {\bf p}_X)^2},  & \text{if } p_X^z > 0 \end{array} \right.
 \end{eqnarray}
Next, deform the contour so that the integration contour is infinitesimally above the real axis:
\begin{eqnarray}
\mathcal{A}(p_X) &=& -ig_S g_D (2^{3/2} \sqrt{\pi} \Delta x)^{3/2}  F_S(p_X, Mv^\mu)  \int_\tau^{\tau+T} dt_1 ~e^{it_1\Delta_D} \int_m^\infty \frac{dE_p}{2(2\pi)^3} \left[\frac{e^{-it_1 E_{ p} } - e^{-i(M - E_X)t_1}}{E_{ p} + E_X - M + i \varepsilon} \right]   \nonumber \\
&& \hspace{1in} \times \left( \frac{2\pi}{iL} \right) \times \left\{ \begin{array}{ll} e^{iL\sqrt{E_p^2 - m^2} - \Delta x^2(\sqrt{E_p^2 - m^2} \hat{z} + {\bf p}_X)^2}, & \text{if } p_X^z < 0  \\ -e^{-iL\sqrt{E_p^2 - m^2} - \Delta x^2(\sqrt{E_p^2 - m^2} \hat{z} - {\bf p}_X)^2},  & \text{if } p_X^z > 0 \end{array} \right.~~.
 \end{eqnarray}
%

\begin{figure}[h]
\includegraphics[width=6cm]{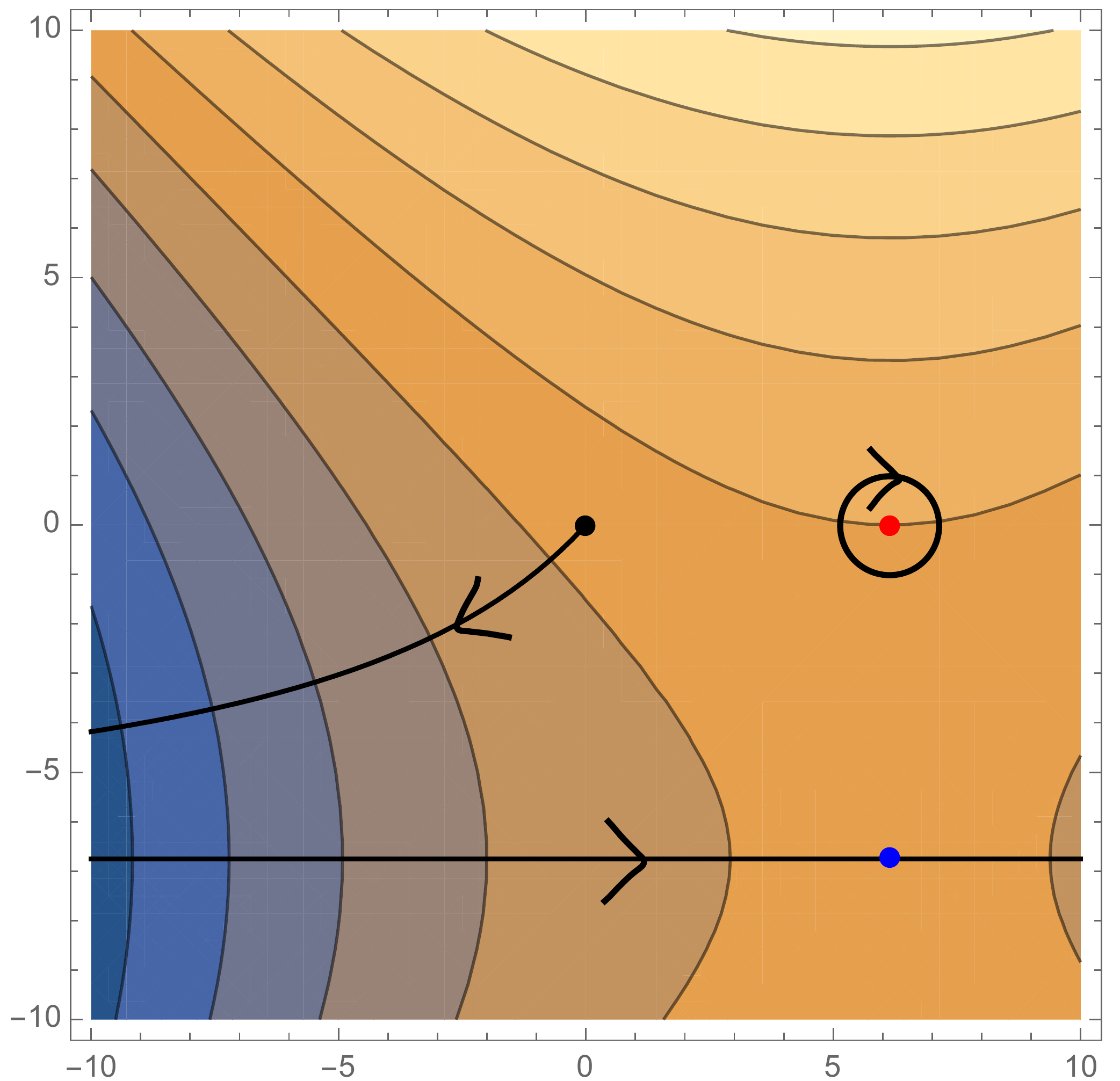}
\caption{The deformed contour when $a>0$.  The red dot is the location of the pole, the blue dot is the location of the saddle point, and the black dot is the location of the integration endpoint.   The colored contour is $\phi(u,v)$, where more blue is more negative, more orange/yellow is more positive.  Ignore the units of the axes.  }
\label{fig:deformedcontour}
\end{figure}

\footnotesize
\noindent
{\bf Aside \# 2}.  Consider the following integral:
\begin{equation}
I = \int_0^\infty dE~ \frac{e^{-\Delta x^2(E-p_0)^2} e^{-iaE}}{E-p_0 + i \varepsilon}
\end{equation}
where $\Delta x$, $p_0$ and $a$ are all positive real numbers, and $\varepsilon$ is positive and infinitesimally small.  The integrand has a pole at $E = p_0 - i\varepsilon$, and a saddle point at $E = p_0 - ia/(2\Delta x^2)$.  The goal is to estimate the value of this integral in the limit that $a\rightarrow\infty$ and $\Delta x \rightarrow \infty$. 
Assuming analyticity, the argument of the exponent can be expressed as the following, letting $E = u + i v$:
\begin{equation}
-\Delta x^2(E-p_0)^2 -iaE \rightarrow \phi(u,v) + i \psi(u,v)
\end{equation}
where
\begin{eqnarray}
\phi(u,v) &=& av + (v^2-(p_0 -u)^2)\Delta x^2 \\
\psi(u,v) &=& -au + 2v(p_0-u)\Delta x^2~~.
\end{eqnarray}
Deform the contour away from the positive real axis, going around the pole.  Choose a contour $C$ such that the segments are along paths of constant $\psi(u,v)$, and one of the path segments is through the saddle point, as shown in Fig.~\ref{fig:deformedcontour}. 
Then, we have
\begin{equation}
I = \int_C dE~ \frac{e^{-\Delta x^2(E-p_0)} e^{-iaE}}{E-p_0 + i \varepsilon} - 2\pi i ~e^{-iap_0}~~.
\end{equation}
There are two contributions to the integral along $C$: the location of the saddle point and the location of the integration endpoint.  The contribution from the saddle point is
\begin{eqnarray}
\int_\text{saddle point} dE~ \frac{e^{-\Delta x^2(E-p_0)} e^{-iaE}}{E-p_0 + i \varepsilon}  
&\sim& \frac{e^{-iap_0}e^{-a^2/(4\Delta x^2)}}{-ia/(2\Delta x^2)} \int_{-\infty}^\infty du~ e^{-u^2\Delta x^2} \\
&=& 2i\sqrt{\pi} \left( \frac{\Delta x}{a}\right) e^{-iap_0}e^{-a^2/(4\Delta x^2)}~~.
\end{eqnarray}
The contribution from the integration endpoint is
\begin{eqnarray}
\int_\text{endpoint} dE~ \frac{e^{-\Delta x^2(E-p_0)} e^{-iaE}}{E-p_0 + i \varepsilon}  &\sim& -\frac{e^{-\Delta x^2 p_0^2}}{p_0} \int_0^{-\infty} dv ~e^{v\left(a + \frac{4p_0^2 \Delta x^4}{a} \right)} \\
&=& \frac{e^{-\Delta x^2 p_0^2}}{ap_0} \left( \frac{1}{1+ \frac{4p_0^2 \Delta x^4}{a^2}} \right)~~.
\end{eqnarray}
To leading order in $1/a$ and $1/\Delta x$,
\begin{eqnarray}
I &\sim& - 2\pi i ~e^{-iap_0} + \frac{e^{-\Delta x^2 p_0^2}}{ap_0} \left( \frac{1}{1+ \frac{4p_0^2 \Delta x^4}{a^2}} \right) + 2i\sqrt{\pi} \left( \frac{\Delta x}{a}\right) e^{-iap_0}e^{-a^2/(4\Delta x^2)}  \\
&=& - 2\pi i ~e^{-iap_0} + \mathcal{O} \left(\frac{e^{-\Delta x^2 p_0^2}}{ap_0}, \frac{ p_0^2 \Delta x^4}{a^2}, \frac{\Delta x}{a} e^{-a^2/\Delta x^2} \right)~~.
\end{eqnarray}
\noindent
If, on the other hand, $a<0$, then one can use the same procedure as above, where again the contour has a contribution from the saddle point in the upper half plane and the integration endpoint, but there is no contribution from the pole, since the original contour can be analytically deformed to the upper-half plane:
\begin{equation}
I \sim \mathcal{O} \left(\frac{e^{-\Delta x^2 p_0^2}}{ap_0}, \frac{ p_0^2 \Delta x^4}{a^2}, \frac{\Delta x}{a} e^{-a^2/\Delta x^2} \right)~~.
\end{equation}


\normalsize
\noindent
The result from Aside \#2 implies that the pole is the dominant contribution to the integral, in the limit that $\Delta x/L \ll 1$, which in turn implies that $p_X^z < 0$, when $t_1 - L$ is positive (the detector is on completely inside the source's lightcone) yielding:
\begin{eqnarray}
\mathcal{A}(p_X) &=& -ig_S g_D (2^{3/2} \sqrt{\pi} \Delta x)^{3/2}  F_S(p_X, Mv^\mu)  \int_\tau^{\tau+T} dt_1 ~e^{it_1(\Delta_D - M + E_X)}  \cdot \frac{1}{2(2\pi)^3} \nonumber \\
&& \hspace{1in} \times \left( \frac{2\pi}{iL} \right) \times (-2\pi i)e^{iL\sqrt{(M-E_X)^2 - m^2} - \Delta x^2(\sqrt{(M-E_X)^2 - m^2} \hat{z} + {\bf p}_X)^2} \nonumber \\
&& \hspace{1.5in}+ ~\mathcal{O}\left(\frac{e^{-\Delta x^2 (M-E_X)^2}}{L(M-E_X)}, \frac{ (M-E_X)^2 \Delta x^4}{L^2}, \frac{\Delta x}{L} e^{-L^2/\Delta x^2} \right)~.
 \end{eqnarray}
Doing the $t_1$ integral:
\begin{eqnarray}
\mathcal{A}(p_X) &\simeq& -ig_S g_D (2^{3/2} \sqrt{\pi} \Delta x)^{3/2}  F_S(p_X, Mv^\mu) \cdot \frac{1}{2(2\pi)^3} \cdot e^{i(\tau + T/2)(\Delta_D - M + E_X)} \frac{2\sin\left((\Delta_D - M + E_X)T/2 \right)}{(\Delta_D - M + E_X)} \nonumber \\
&& \hspace{1in} \times \left( \frac{2\pi}{iL} \right) \times  (-2\pi i)e^{iL\sqrt{(M-E_X)^2 - m^2} - \Delta x^2(\sqrt{(M-E_X)^2 - m^2} \hat{z} + {\bf p}_X)^2} ~.
\end{eqnarray}
Squaring the amplitude, summing over the detector system, and taking there to be only one particle in the $X$ state:
\begin{eqnarray}
\mathcal{P} &=& \int \frac{d^3{\bf p}_X}{(2\pi)^3}  \int d\Delta_D~ \rho(\Delta_D) ~ \Bigg| \sum_a |U_a|^2 \mathcal{A}(p_X)\Bigg|^2~.
\end{eqnarray}
Note that the final-state phase space
measure lacks the factor of $1/2E_X$, since we used non-relativistic normalization 
when we defined in source's wavepacket in Appendix~\ref{intialstateofsource}.  
Using $\sin^2(xT/2)/x^2 \simeq \frac{T}{2}\pi \delta(x) + \mathcal{O}(1/T)$, do the $\Delta_D$ integral, pulling out the form factor $F_S$, since it is the approximately the same for all neutrino species: 
\begin{eqnarray}
\mathcal{P} &=& \mathcal{C} \int \frac{d^3{\bf p}_X}{(2\pi)^3} \rho(M-E_X) |F_S(p_X)|^2 \Bigg| \sum_a |U_a|^2 e^{iL\sqrt{(M-E_X)^2 - m_a^2} - \Delta x^2\left(\sqrt{(M-E_X)^2 - m_a^2} \hat{z} + {\bf p}_X\right)^2} \Bigg|^2
\end{eqnarray}
where 
\begin{equation}
\mathcal{C} \equiv  \frac{(2^{3/2} \sqrt{\pi } \Delta x)^3 g_S^2 g_D^2 T }{8\pi L^2} 
\end{equation}
Let $E_X = \sqrt{{\bf p}_X^2 + M_X^2}$.  In the limit that $m=0$, then the $\Delta x$ Gaussian is centered at ${\bf p}_X^* = -\left(\frac{M^2 - M_X^2}{2M}\right)\hat{z}$.  

In the limit that the support of the $\Delta x$ Gaussian is much smaller than the scale of variations of the rest of the integrand,  we can expand the argument of the Gaussian around ${\bf p}_X^* = -\left(\frac{M^2 - M_X^2}{2M}\right)\hat{z}$, as well as expanding in $m$:
\small
\begin{eqnarray} 
-2\Delta x^2\left(\hat{z}\sqrt{(M-E_X)^2 - m_a^2} + {\bf p}_X \right)^2 &\simeq& -8\Delta x^2\left[ \frac{M^4}{(M^2+M_X^2)^2} + \frac{2 m_a^2 M^4(M^2+2M_X^2)}{(M^2 - M_X^2)(M^2 + M_X^2)^3} + \mathcal{O}(m_a^4) \right]  \left( p^z_X - {\bf p}_X^*\cdot \hat{z}\right)^2 \nonumber \\
&& \hspace{0.5in} -~8\Delta x^2 \left[ \frac{m_a^2M^3 }{M^4-M_X^4} + \mathcal{O}(m_a^4) \right]\left( p^z_X - {\bf p}_X^*\cdot \hat{z}\right) \nonumber \\
&& \hspace{0.5in} -~2\Delta x^2 \left[  (p_X^x)^2 + (p_X^y)^2 \right]
\end{eqnarray}
\normalsize
If the neutrino is ultra-relativistic, the terms proportional to $m^2$ are very subdominant.  So, to a very good approximation, all neutrino species have the same saddle point.  Doing the momentum integrals with saddle point methods (here again we assume that the form factor is slowly varying, $ \Lambda_\text{QCD}^{-1} \gg \Delta x, L$):
\begin{eqnarray}
\mathcal{P} &\sim& \mathcal{C} \frac{1}{(2\pi)^3}  \rho(M-E_X^*) |F_S(p_X^*)|^2 \Bigg| \sum_a |U_a|^2 e^{iL\sqrt{(M-E^*_X)^2 - m_a^2}} \Bigg|^2\nonumber \\
&& \hspace{1in} \times ~ \int_{-\infty}^\infty dp_x~dp_y~dp_z  e^{ -8\Delta x^2\left[ \frac{M^4}{(M^2+M_X^2)^2}  \right]  \left( p_z- {\bf p}_X^*\cdot \hat{z}\right)^2 - 2\Delta x^2(p_x^2 + p_y^2)}  \\
&=& \mathcal{C} \frac{1}{(2\pi)^3} \rho(M-E_X^*) |F_S(p_X^*)|^2 \Bigg| \sum_a |U_a|^2 e^{iL\sqrt{(M-E^*_X)^2 - m_a^2}} \Bigg|^2\nonumber \\
&& \hspace{0.2in} \times ~ \left(\sqrt{\frac{\pi}{2\Delta x^2}}\right)^2 \sqrt{\frac{\pi}{\frac{8\Delta x^2M^4}{(M^2+M_X^2)^2}}}
\end{eqnarray}
where $E_X^* \equiv \sqrt{|{\bf p}_X^*|^2 + M_X^2}$.  Inputting the expression for $\mathcal{C}$, we have:
\begin{equation}
\mathcal{P} = \frac{g_S^2 g_D^2 T }{8 \pi L^2} \frac{(M^2 + M_X^2)}{2M^2} \rho(M-E_X^*) |F_S(p_X^*)|^2 \Bigg| \sum_a |U_a|^2 e^{iL\sqrt{(M-E^*_X)^2 - m_a^2}} \Bigg|^2
\end{equation}
Here, $M - E_X^* = (M^2-M_X^2)/(2M)$, which can be identified as the energy of the neutrino, if it were massless.  
If $M-M_X = \Delta$, and $\Delta/M \ll 1$, then $M-E_X^* = \Delta + \mathcal{O}(\Delta^2)$, and we have
\begin{equation}
\label{eq:finally}
\mathcal{P} = \frac{g_S^2 g_D^2 T }{8 \pi L^2} \rho(\Delta) |F_S(p_X^*)|^2 \Bigg| \sum_a |U_a|^2 e^{iL\sqrt{\Delta^2 - m_a^2}} \Bigg|^2 + \mathcal{O} \left( \frac{\Delta}{M} \right)
\end{equation}
The conditions for this conclusion to be valid are that $L \gg \Delta x \gg \Lambda_\text{QCD}^{-1}$, using $\Lambda_\text{QCD}$ as the typical scale of variation of the form factor.
If the form factor $F_S = 1$, and $g_D = g_S$, then we recover the result in Eq.~\eqref{simpleprob} 
in Section~\ref{stationarysource}.

\bibliographystyle{JHEP}

\bibliography{bib}{}

\providecommand{\href}[2]{#2}\begingroup\raggedright\begin{thebibliography}{10}

\bibitem{Esteban:2016qun}
I.~Esteban, M.~C. Gonzalez-Garcia, M.~Maltoni, I.~Martinez-Soler and
  T.~Schwetz, \emph{{Updated fit to three neutrino mixing: exploring the
  accelerator-reactor complementarity}},
  \href{https://doi.org/10.1007/JHEP01(2017)087}{\emph{JHEP} {\bfseries 01}
  (2017) 087}, [\href{https://arxiv.org/abs/1611.01514}{{\ttfamily
  1611.01514}}].

\bibitem{deSalas:2017kay}
P.~F. de~Salas, D.~V. Forero, C.~A. Ternes, M.~Tortola and J.~W.~F. Valle,
  \emph{{Status of neutrino oscillations 2017}},
  \href{https://arxiv.org/abs/1708.01186}{{\ttfamily 1708.01186}}.

\bibitem{Patrignani:2016xqp}
{\scshape Particle Data Group} collaboration, C.~Patrignani et~al.,
  \emph{{Review of Particle Physics}},
  \href{https://doi.org/10.1088/1674-1137/40/10/100001}{\emph{Chin. Phys.}
  {\bfseries C40} (2016) 100001}.

\bibitem{Nussinov:1976uw}
S.~Nussinov, \emph{{Solar Neutrinos and Neutrino Mixing}},
  \href{https://doi.org/10.1016/0370-2693(76)90648-1}{\emph{Phys. Lett.}
  {\bfseries 63B} (1976) 201--203}.

\bibitem{Kayser:1981ye}
B.~Kayser, \emph{{On the Quantum Mechanics of Neutrino Oscillation}},
  \href{https://doi.org/10.1103/PhysRevD.24.110}{\emph{Phys. Rev.} {\bfseries
  D24} (1981) 110}.

\bibitem{Giunti:1991sx}
C.~Giunti, C.~W. Kim and U.~W. Lee, \emph{{Coherence of neutrino oscillations
  in vacuum and matter in the wave packet treatment}},
  \href{https://doi.org/10.1016/0370-2693(92)90308-Q}{\emph{Phys. Lett.}
  {\bfseries B274} (1992) 87--94}.

\bibitem{Giunti:2002xg}
C.~Giunti, \emph{{Neutrino wave packets in quantum field theory}},
  \href{https://doi.org/10.1088/1126-6708/2002/11/017}{\emph{JHEP} {\bfseries
  11} (2002) 017}, [\href{https://arxiv.org/abs/hep-ph/0205014}{{\ttfamily
  hep-ph/0205014}}].

\bibitem{Giunti:1993se}
C.~Giunti, C.~W. Kim, J.~A. Lee and U.~W. Lee, \emph{{On the treatment of
  neutrino oscillations without resort to weak eigenstates}},
  \href{https://doi.org/10.1103/PhysRevD.48.4310}{\emph{Phys. Rev.} {\bfseries
  D48} (1993) 4310--4317},
  [\href{https://arxiv.org/abs/hep-ph/9305276}{{\ttfamily hep-ph/9305276}}].

\bibitem{Grimus:1996av}
W.~Grimus and P.~Stockinger, \emph{{Real oscillations of virtual neutrinos}},
  \href{https://doi.org/10.1103/PhysRevD.54.3414}{\emph{Phys. Rev.} {\bfseries
  D54} (1996) 3414--3419},
  [\href{https://arxiv.org/abs/hep-ph/9603430}{{\ttfamily hep-ph/9603430}}].

\bibitem{Kiers:1995zj}
K.~Kiers, S.~Nussinov and N.~Weiss, \emph{{Coherence effects in neutrino
  oscillations}}, \href{https://doi.org/10.1103/PhysRevD.53.537}{\emph{Phys.
  Rev.} {\bfseries D53} (1996) 537--547},
  [\href{https://arxiv.org/abs/hep-ph/9506271}{{\ttfamily hep-ph/9506271}}].

\bibitem{Giunti:1997wq}
C.~Giunti and C.~W. Kim, \emph{{Coherence of neutrino oscillations in the wave
  packet approach}},
  \href{https://doi.org/10.1103/PhysRevD.58.017301}{\emph{Phys. Rev.}
  {\bfseries D58} (1998) 017301},
  [\href{https://arxiv.org/abs/hep-ph/9711363}{{\ttfamily hep-ph/9711363}}].

\bibitem{Stockinger:2000sk}
P.~Stockinger, \emph{{Introduction to a field-theoretical treatment of neutrino
  oscillations}},
  \href{https://doi.org/10.1007/s12043-000-0017-1}{\emph{Pramana} {\bfseries
  54} (2000) 203--214}.

\bibitem{Beuthe:2001rc}
M.~Beuthe, \emph{{Oscillations of neutrinos and mesons in quantum field
  theory}}, \href{https://doi.org/10.1016/S0370-1573(02)00538-0}{\emph{Phys.
  Rept.} {\bfseries 375} (2003) 105--218},
  [\href{https://arxiv.org/abs/hep-ph/0109119}{{\ttfamily hep-ph/0109119}}].

\bibitem{Beuthe:2002ej}
M.~Beuthe, \emph{{Towards a unique formula for neutrino oscillations in
  vacuum}}, \href{https://doi.org/10.1103/PhysRevD.66.013003}{\emph{Phys. Rev.}
  {\bfseries D66} (2002) 013003},
  [\href{https://arxiv.org/abs/hep-ph/0202068}{{\ttfamily hep-ph/0202068}}].

\bibitem{Giunti:2003ax}
C.~Giunti, \emph{{Coherence and wave packets in neutrino oscillations}},
  \href{https://doi.org/10.1023/B:FOPL.0000019651.53280.31}{\emph{Found. Phys.
  Lett.} {\bfseries 17} (2004) 103--124},
  [\href{https://arxiv.org/abs/hep-ph/0302026}{{\ttfamily hep-ph/0302026}}].

\bibitem{Bernardini:2004sw}
A.~E. Bernardini and S.~De~Leo, \emph{{An Analytic approach to the wave packet
  formalism in oscillation phenomena}},
  \href{https://doi.org/10.1103/PhysRevD.70.053010}{\emph{Phys. Rev.}
  {\bfseries D70} (2004) 053010},
  [\href{https://arxiv.org/abs/hep-ph/0411134}{{\ttfamily hep-ph/0411134}}].

\bibitem{Bernardini:2006ak}
A.~E. Bernardini, M.~M. Guzzo and F.~R. Torres, \emph{{Second-order corrections
  to neutrino two-flavor oscillation parameters in the wave packet approach}},
  \href{https://doi.org/10.1140/epjc/s10052-006-0032-6}{\emph{Eur. Phys. J.}
  {\bfseries C48} (2006) 613},
  [\href{https://arxiv.org/abs/hep-ph/0612001}{{\ttfamily hep-ph/0612001}}].

\bibitem{Akhmedov:2008jn}
E.~K. Akhmedov, J.~Kopp and M.~Lindner, \emph{{Oscillations of Mossbauer
  neutrinos}}, \href{https://doi.org/10.1088/1126-6708/2008/05/005}{\emph{JHEP}
  {\bfseries 05} (2008) 005},
  [\href{https://arxiv.org/abs/0802.2513}{{\ttfamily 0802.2513}}].

\bibitem{Cohen:2008qb}
A.~G. Cohen, S.~L. Glashow and Z.~Ligeti, \emph{{Disentangling Neutrino
  Oscillations}},
  \href{https://doi.org/10.1016/j.physletb.2009.06.020}{\emph{Phys. Lett.}
  {\bfseries B678} (2009) 191--196},
  [\href{https://arxiv.org/abs/0810.4602}{{\ttfamily 0810.4602}}].

\bibitem{Akhmedov:2009rb}
E.~K. Akhmedov and A.~{\relax Yu}. Smirnov, \emph{{Paradoxes of neutrino
  oscillations}}, \href{https://doi.org/10.1134/S1063778809080122}{\emph{Phys.
  Atom. Nucl.} {\bfseries 72} (2009) 1363--1381},
  [\href{https://arxiv.org/abs/0905.1903}{{\ttfamily 0905.1903}}].

\bibitem{Naumov:2010um}
D.~V. Naumov and V.~A. Naumov, \emph{{A Diagrammatic treatment of neutrino
  oscillations}},
  \href{https://doi.org/10.1088/0954-3899/37/10/105014}{\emph{J. Phys.}
  {\bfseries G37} (2010) 105014},
  [\href{https://arxiv.org/abs/1008.0306}{{\ttfamily 1008.0306}}].

\bibitem{Akhmedov:2010ms}
E.~K. Akhmedov and J.~Kopp, \emph{{Neutrino oscillations: Quantum mechanics vs.
  quantum field theory}}, \href{https://doi.org/10.1007/JHEP04(2010)008,
  10.1007/JHEP10(2013)052}{\emph{JHEP} {\bfseries 04} (2010) 008},
  [\href{https://arxiv.org/abs/1001.4815}{{\ttfamily 1001.4815}}].

\bibitem{Akhmedov:2012uu}
E.~Akhmedov, D.~Hernandez and A.~Smirnov, \emph{{Neutrino production coherence
  and oscillation experiments}},
  \href{https://doi.org/10.1007/JHEP04(2012)052}{\emph{JHEP} {\bfseries 04}
  (2012) 052}, [\href{https://arxiv.org/abs/1201.4128}{{\ttfamily 1201.4128}}].

\bibitem{Jones:2014sfa}
B.~J.~P. Jones, \emph{{Dynamical pion collapse and the coherence of
  conventional neutrino beams}},
  \href{https://doi.org/10.1103/PhysRevD.91.053002}{\emph{Phys. Rev.}
  {\bfseries D91} (2015) 053002},
  [\href{https://arxiv.org/abs/1412.2264}{{\ttfamily 1412.2264}}].

\bibitem{An:2016pvi}
{\scshape Daya Bay} collaboration, F.~P. An et~al., \emph{{Study of the wave
  packet treatment of neutrino oscillation at Daya Bay}},
  \href{https://doi.org/10.1140/epjc/s10052-017-4970-y}{\emph{Eur. Phys. J.}
  {\bfseries C77} (2017) 606},
  [\href{https://arxiv.org/abs/1608.01661}{{\ttfamily 1608.01661}}].

\bibitem{Boyanovsky:2011xq}
D.~Boyanovsky, \emph{{Short baseline neutrino oscillations: when entanglement
  suppresses coherence}},
  \href{https://doi.org/10.1103/PhysRevD.84.065001}{\emph{Phys. Rev.}
  {\bfseries D84} (2011) 065001},
  [\href{https://arxiv.org/abs/1106.6248}{{\ttfamily 1106.6248}}].

\bibitem{Lello:2012gi}
L.~Lello and D.~Boyanovsky, \emph{{Searching for sterile neutrinos from $\pi$
  and $K$ decays}},
  \href{https://doi.org/10.1103/PhysRevD.87.073017}{\emph{Phys. Rev.}
  {\bfseries D87} (2013) 073017},
  [\href{https://arxiv.org/abs/1208.5559}{{\ttfamily 1208.5559}}].

\bibitem{Jenkins:1990jv}
E.~E. Jenkins and A.~V. Manohar, \emph{{Baryon chiral perturbation theory using
  a heavy fermion Lagrangian}},
  \href{https://doi.org/10.1016/0370-2693(91)90266-S}{\emph{Phys. Lett.}
  {\bfseries B255} (1991) 558--562}.

\bibitem{Schwinger:1951nm}
J.~S. Schwinger, \emph{{On gauge invariance and vacuum polarization}},
  \href{https://doi.org/10.1103/PhysRev.82.664}{\emph{Phys. Rev.} {\bfseries
  82} (1951) 664--679}.

\bibitem{Peskin:1983up}
M.~E. Peskin, \emph{{ASPECTS OF THE DYNAMICS OF HEAVY QUARK SYSTEMS}},  in
  \emph{{Dynamics and spectroscopy at high-energy: Proceedings, 11th SLAC
  summer institute on particle physics (SSI 83), Stanford, Calif., 18-29 Jul
  1983}}, 1983.

\bibitem{Manohar:2000dt}
A.~V. Manohar and M.~B. Wise, \emph{{Heavy quark physics}}, {\emph{Camb.
  Monogr. Part. Phys. Nucl. Phys. Cosmol.} {\bfseries 10} (2000) 1--191}.

\bibitem{sakurai1995}
J.~J. Sakurai, \emph{Modern quantum mechanics, revised edition}.
\newblock Addison Wesley, 1995.

\bibitem{Chiu:1977ds}
C.~B. Chiu, E.~C.~G. Sudarshan and B.~Misra, \emph{{Time Evolution of Unstable
  Quantum States and a Resolution of Zeno's Paradox}},
  \href{https://doi.org/10.1103/PhysRevD.16.520}{\emph{Phys. Rev.} {\bfseries
  D16} (1977) 520--529}.

\bibitem{Dickinson:2016oiy}
R.~Dickinson, J.~Forshaw and P.~Millington, \emph{{Probabilities and signalling
  in quantum field theory}},
  \href{https://doi.org/10.1103/PhysRevD.93.065054}{\emph{Phys. Rev.}
  {\bfseries D93} (2016) 065054},
  [\href{https://arxiv.org/abs/1601.07784}{{\ttfamily 1601.07784}}].

\bibitem{fermi1932}
E.~Fermi, \emph{Quantum theory of radiation}, {\emph{Reviews of modern physics}
  {\bfseries 4} (1932) 87}.

\bibitem{Fleischhauer}
M.~Fleischhauer, \emph{Quantum-theory of photodetection without the rotating
  wave approximation}, {\emph{J Phys. A:~Math Gen.} {\bfseries 31} (1998) 453}.

\bibitem{Mandel:1995seg}
L.~Mandel and E.~Wolf, \emph{{Optical Coherence and Quantum Optics}}.
\newblock Cambridge Univ. Pr., 1995.

\bibitem{Loudon:2000}
R.~Loudon, \emph{{The Quantum Theory of Light}}.
\newblock Oxford Science Pub., 2000.

\bibitem{Shifman:1987rj}
M.~A. Shifman and M.~B. Voloshin, \emph{{On Production of d and D* Mesons in B
  Meson Decays}}, {\emph{Sov. J. Nucl. Phys.} {\bfseries 47} (1988) 511}.

\end{thebibliography}\endgroup

\end{document}